\documentclass[final,5p,times,twocolumn]{elsarticle}
\usepackage{amsmath,amsfonts,amssymb,amstext}
\usepackage{psfrag}
\usepackage{graphicx,color} 
\usepackage{subfig} 
\usepackage{tabularx}
\usepackage[siunitx]{circuitikz}
\usepackage{placeins}
\usepackage{booktabs, multicol, multirow}
\usepackage[utf8]{inputenc}
\usepackage[hidelinks]{hyperref}
\usepackage[hyphenbreaks]{breakurl}

\journal{International Journal of Electrical Power \& Energy Systems}

\usepackage{acronym}
\newacro{PLC}{power line communication}
\newacro{OFDM}{orthogonal frequency-division multiplexing}
\newacro{HS-OFDM}{Hermitian symmetric OFDM}
\newacro{PSD}{power spectral density}
\newacro{nSNR}{normalized signal to noise ratio}
\newacroindefinite{nSNR}{an}{a}
\newacro{CFR}{channel frequency response}
\newacro{LTV}{linear time-variant}
\newacro{LPTV}{linear periodically time variant}
\newacroindefinite{LPTV}{an}{a}
\newacro{LTI}{linear time-invariant}
\newacroindefinite{LTI}{an}{a}
\newacro{DFT}{discrete Fourier transform}
\newacro{AWGN}{additive white Gaussian noise}
\newacro{SNR}{signal-to-noise ratio}
\newacroindefinite{SNR}{an}{a}
\newacro{FT}{Fourier transform}
\newacro{CIR}{channel impulse response}
\newacro{DMT}{discrete multitone modulation}
\newacro{ISI}{intersymbol interference}
\newacro{HV}{high voltage}
\newacro{MV}{medium voltage}
\newacro{LV}{low voltage}
\newacro{TEM}{transverse electromagnetic}
\newacro{SG}{smart grid}
\newacro{IDFT}{inverse discrete Fourier transform}
\newacro{IFT}{inverse Fourier transform}
\newacro{FDR}{frequency-domain reflectometry}
\newacro{PSLR}{peak-to-sidelobe level ratio}
\newacro{ISLR}{integrated-sidelobe level ratio}
\newacro{UWB}{ultra-wideband}
\newacro{CSS}{chirp spread spectrum}
\newacro{BPSK}{binary phase-shift keying}
\newacro{NB}{narrowband}
\newacro{BB}{broadband}
\newacro{EMC}{electromagnetic compatibility}
\newacro{PCR}{pulse compression ratio}
\newacro{PCG}{pulse compression gain}
\newacro{CENELEC}{European Committee for Electrotechnical Standardization (\textit{Comit\'e Europ\'en de Normalisation \'Electrotechnique} in French)}
\newacro{ARIB}{Association of Radio Industries and Businesses}
\newacro{FCC}{Federal Communications Commission}
\newacro{IoT}{Internet of Things}
\newacro{SG}{Smart Grid}
\newacro{AMR}{automatic meter reading}
\newacro{AMI}{advanced metering	infrastructure}
\newacro{LIF}{low-impedance fault}
\newacro{HIF}{high-impedance fault}
\newacro{FDR}{frequency-domain reflectometry}
\newacro{SFR}{swept-frequency reflectometry}
\newacro{TDR}{time-domain reflectometry}
\newacro{JTFDR}{joint time-frequency domain reflectometry}
\newacro{WSS}{wide-sense stationary}
\newacro{LPTV}{linear periodically time-variant}
\newacro{LTI}{linear time-invariant}
\newacro{PAPR}{peak-to-average power ratio}
\newacro{PLM}{power line modem}

\usepackage{float}
\newcommand{\Tau}{\mathrm{T}}

\usepackage{url}

\usepackage{flushend}

\begin{document}

\begin{frontmatter}
	
\title{On the Suitability of PLC Pulses for Power Line Fault Sensing\\ via Time-Domain Reflectometry}
 
	
%

\author{{Lucas Giroto de Oliveira$^{1}$, Mateus de L. Filomeno$^{1}$, Luiz Fernando Colla$^{2}$,\\
		H. Vincent Poor$^{3}$, and Mois\'es V. Ribeiro$^{1,4}$}}
	
\address{$^1$ Electrical Engineering Department\\	
	Federal University of Juiz de Fora\\
    Juiz de Fora, Brazil\\
	$^2$ Copel Distribui\c{c}\~ao LTD\\	
	Curitiba, Brazil\\
	$^3$ Department of Electrical Engineering\\
	Princeton University\\
	Princeton, USA\\
    $^4$ Smarti9 LTD.\\
	Juiz de Fora, Brazil\\
	E-mail: lgiroto@ieee.org, mateus.lima@engenharia.ufjf.br, luiz.colla@copel.com,\\ poor@princeton.edu, mribeiro@ieee.org}

\begin{abstract}
This work discusses the suitability of typical power line communication (PLC) pulses for fault sensing in power lines via pulse-compression time-domain reflectometry (TDR). For this purpose, we first carefully outline a TDR system operating over a power distribution network, discussing its limitations and ultimately resorting to the pulse compression procedure. Next, we present closed-form expressions for pulses and their autocorrelation functions of typical PLC modulation schemes, namely Hermitian symmetric orthogonal frequency-division multiplexing (HS-OFDM), impulsive ultra-wideband (UWB), and chirp spread spectrum (CSS). Furthermore, different metrics are used on the provided expressions in order to evaluate the suitability of the considered PLC pulses for providing proper TDR measurements, i.e., \textit{reflectograms}, in terms of resolution, effectiveness of the pulse compression, distortion, and range. Considering the scenarios of European underground low-voltage and US overhead medium-voltage power distribution networks, we finally carry out a comparative numerical analysis among the considered PLC pulses in terms of the aforementioned metrics for frequency bands comprising narrowband- and broadband-PLC, with a highlight on compliance to regulatory constraints. Based on the achieved results, we show that the use of UWB pulses allows the obtaining of a larger number of reflectograms in a given time interval, while HS-OFDM and CSS pulses provide higher reflectogram quality and better resolution for a given occupied frequency bandwidth. Also, we show that NB-PLC pulses are suitable for most distribution network scenarios, being the use of BB-PLC left for cases where very fine resolutions are desired.
\end{abstract}

%


\end{frontmatter}

\section{Introduction}\label{sec:introduction}


In the modern society, electrical energy is a paramount resource to both industry and residential consumers. Thus, the almost ubiquitous power transmission and distribution networks have been growing in size and complexity. The efficiency of such networks depends partly on their topology and partly on the physical characteristics and integrity of their components. The components of these networks are subjects to faults, which must be promptly handled in order to avoid interruptions in the power supply and consequent economic losses, besides avoiding public hazard.

Among the different types of faults in these networks are termination impedance changes, distributed faults and lumped faults \cite{passerini2018_1}. The first are caused by either natural behavior of the loads or faults near the load, while the second are result of processes such as aging and degradation of cables. Lumped or localized faults, in their turn, are result of break of cables, as well as contact with threes or animals. The described factors generate \acp{LIF} and \acp{HIF}, being the former easily detected by conventional protection schemes. The detection and location of \acp{HIF}, on the other hand, requires more careful approaches. Out of these reasons, fault sensing in power lines has long been subject of interest of both industry and academia \cite{sedighizadeh2010,ghaderi2017}. In this context, the application of traveling waves, widely used for fault sensing in wired networks, constitutes an attractive approach which has been gaining attention for detection and location of impedance discontinuities caused by elements such as branches, loads and faults \cite{auzanneau2013}. For this purpose, wired networks are handled as transmission lines through which \ac{TEM} waves travel \cite{paul2007}.

The use of the concept of traveling waves enables reflectometry techniques, which consist of injecting a pulse into the analyzed network and capturing the raised reflections, which compose a \textit{reflectogram}, at a single measurement point by an instrument called \textit{reflectometer}. As such signal propagates along the network, each impedance discontinuity causes part of the signal energy to be reflected and, through analysis of the reflected signal or \textit{echo}, it is possible to obtain information on the network topology. Among different reflectometry approaches are the \ac{FDR}, also known as \ac{SFR}, and the \ac{TDR} \cite{furse2006}. There is also the \ac{JTFDR}, which benefits from the features from both \ac{FDR} and \ac{TDR} and overcomes some limitations associated to these approaches at the cost of more complex post-processing \cite{wang2010}.

One interesting alternative for enabling fault sensing via reflectometry in power lines is the use of \ac{PLC} technology \cite{costa2017,filomeno2018}, which has proven to be a key technology in the context of \ac{SG} \cite{galli2011,dib2018,massi2018} and also plays an important role in Industry 4.0, as it enables \ac{IoT} applications that can perform industrial automation \cite{wollschlaeger2017,fernandes2018}. The usage of a broad frequency bandwidth by \ac{PLC} technologies empowers approaches based on traveling waves and enables the so-called \textit{broadband reflectometry} \cite{cataldo2011}, which provides reflectograms with finer resolution. In the specific case of power distribution networks, approaches such as \ac{FDR}-based techniques \cite{passerini2017}, techniques that combine impedance and \ac{TDR} measurements \cite{milioudis2015}, as well as the use of \ac{JTFDR} \cite{wang2010,huo2018} have been investigated. The sole use of \ac{TDR} principles for fault sensing in power lines, subject of study in the past \cite{taylor1996,bo1999}, has been recently revisited in the literature \cite{paulis2016}. One of its main advantages is the reduced computational complexity due to simplified post-processing, as the reflectogram is directly obtained from the captured reflections. This characteristic allows on-line obtaining of reflectograms \cite{hassen2015}, efficiently enabling both symbol level sensing and mains level sensing \cite{passerini2018_2}. To the best of the authors' knowledge, however, \ac{TDR}-based techniques have only been deeply investigated to fault sensing in common wired networks and a \ac{PLC}-focused approach has not been presented. For this reason, we turn our attention to fault sensing in power lines via \ac{PLC}-based \ac{TDR}.

In the context of this work, we focus on pulse-compression \ac{TDR} \cite{chang2015,hassen2015}, which consists of convolving the captured reflected signal with a matched filter to the injected pulse. Analyzing the resulting reflectogram, such procedure compresses the originally injected pulse into a narrower pulse, therefore performing resolution improvement and increasing the average transmitted power proportionally to the time duration of the originally injected pulse \cite{chang2015}. In contrast to typical works on reflectometry-based fault sensing in wired networks, we discuss the aspects of resolution, range and quality of the obtained reflectograms by approaching the pulse-compression \ac{TDR} in power lines as a \textit{wired radar}. Given this context, the main contributions of this work are summarized as follows.
\begin{enumerate}
	\item Description of a pulse-compression \ac{TDR} system over a generic power distribution network. Based on this formulation, we present typical \ac{PLC} pulses, i.e., \ac{HS-OFDM} \cite{giroto2018}, first \cite{salmento2017} and second \cite{tonello2011} derivatives of the Gaussian pulse (which belong to the impulsive \ac{UWB} category), and \ac{CSS} \cite{dib_dissert2018}, and provide closed-form expressions for their autocorrelation functions, which result from the pulse-compression procedure.
	\item Comparative analysis between the aforementioned \ac{PLC} pulses, covering typical European underground \ac{LV} and US overhead \ac{MV} scenarios, using range resolution, \ac{PSLR}, \ac{ISLR}, and maximum unambiguous range as performance metrics and highlighting results for \ac{PLC}-based pulse-compression \ac{TDR} systems complying with \ac{NB}- and \ac{BB}-\ac{PLC} regulations.
\end{enumerate}
Our major findings are as follows:
\begin{enumerate}
	\item Within a same time interval, the use of \ac{UWB} pulses results in the obtaining of a much higher number of reflectograms with significantly finer resolution than with the use of \ac{HS-OFDM} and \ac{CSS} pulses with same occupied frequency bandwidth, with the drawback of reduced reflectogram quality due to high sidelobe levels.
	\item Also for a given occupied frequency bandwidth, the best range resolution is achieved by \ac{HS-OFDM} and \ac{CSS} pulses, followed by the second and first derivatives of the Gaussian pulse, in this order.
	\item The use of \ac{HS-OFDM} and \ac{CSS} pulses, although resulting in longer pulses with coarser resolution, results in a more effective pulse compression and in reflectograms with less significant distortion.
	\item NB-PLC pulses provide proper resolution for most distribution network scenarios, being therefore the use of BB-PLC left for cases where very fine resolutions are desired, such as very short distribution network sections.
\end{enumerate}

The remainder of this paper is organized as follows. Section II describes the scenario in which our discussion takes place. Section III deals with the compression of typical \ac{PLC} pulses. Section IV presents performance metrics for assessing the quality of obtained reflectograms. Section V carries out a numerical analysis of case studies. Finally, concluding remarks are placed in Section VI.

\section{Problem Statement}\label{sec:statement}

\begin{figure}[!b]
	\centering	
	\subfloat[ ]{
		\psfrag{A}[c][c]{$Z_0,\gamma,\ell_{0}$}
		\psfrag{B}[c][c]{$Z_0,\gamma,\ell_{M}$}
		\psfrag{C}[c][c]{$Z_{L,0}$}
		\psfrag{D}[c][c]{$Z_{L,M}$}
		\includegraphics[width=8cm]{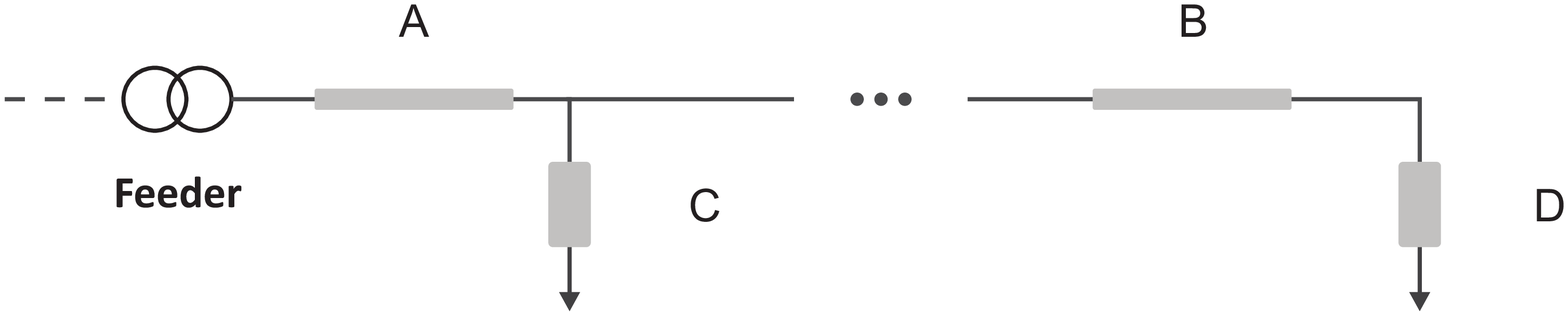}\label{fig:dist-network}
	}\\	
	\subfloat[ ]{
		\psfrag{A}[c][c]{$Z_0,\gamma,\ell_{0}$}
		\psfrag{B}[c][c]{$Z_0,\gamma,\ell_{M}$}
		\psfrag{C}[c][c]{$Z_{L,0}$}
		\psfrag{D}[c][c]{$Z_{L,M}$}
		\psfrag{F}[c][c]{\hspace{0.2cm}$Z_{fault}$}
		\includegraphics[width=8cm]{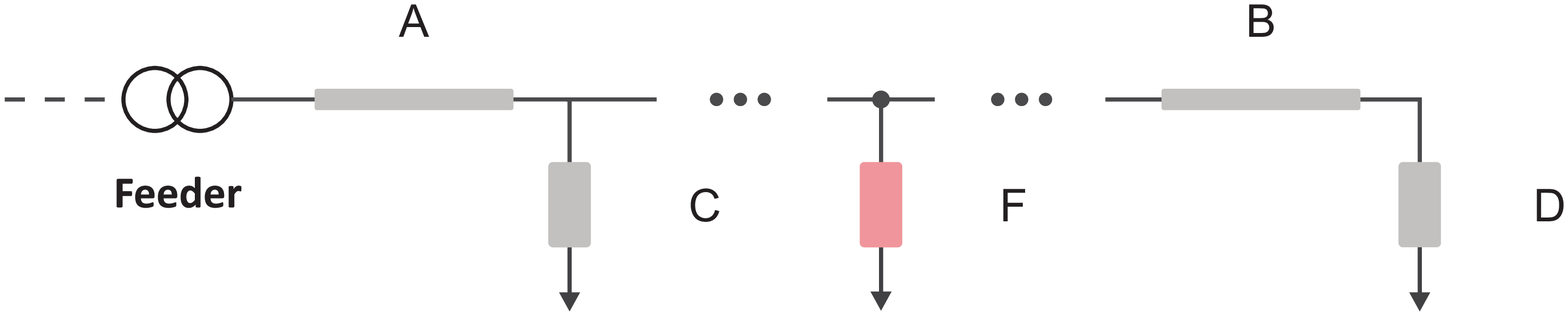}\label{fig:dist-network-fault}
	}\\
	\caption{Sketch of a power distribution network (a) under normal operation and (b) under fault ocurrence.}	
\end{figure}

To start our discussion, let a distribution network be constituted by power supply cable sections and branches ending at loads. From the perspective of transmission line theory, the power supply cable sections are two-conductor transmission line sections with frequency-dependent characteristic impedance $Z_0(f)$ and propagation constant $\gamma(f)$, where $f$ refers to frequency, which are determined by constructive characteristics of the power supply cable. The branches, in their turn, are bridged taps with finite length ending at a load. As an example, we have the sketch of a distribution network in Fig~\ref{fig:dist-network}. In this figure, the distribution network is constituted by a feeder, $M$ line sections, and $M$ branches ending at loads. The $m^{th}$ line section has length $\ell_{m}$ and the $m^{th}$ branch ends at a load with impedance $Z_{L,m}(f)$, where $m=0,...,M$. For the sake of simplicity, the variable $f$ has been omitted in this figure.

\begin{figure}[!t]
	\centering
	\psfrag{h}[c][c]{$h(t)$}
	\includegraphics[width=7.5cm]{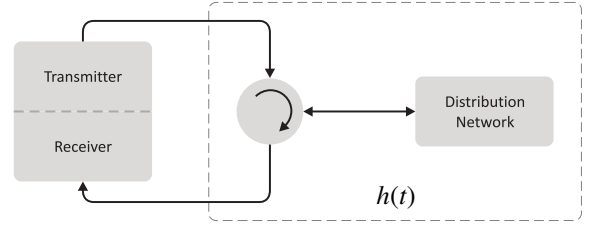}
	\caption{Reflectometry system over a power distribution network with impulse response $h(t)$ of the reflection channel.} \label{fig:h_circulator}
\end{figure}

Let a \ac{PLC} transceiver be connected to a point of that distribution network, experiencing an input impedance $Z_{in}(f)$ that results from the combination of the aforementioned elements~\cite{paul2007}. Such transceiver injects pulses and captures their resulting reflections at a single measurement point as in Figure~\ref{fig:h_circulator}. Also, let us assume the use of an ideal circulator, which completely separates the signals injected by the transmitting part of the \ac{PLM} from the reflections captured by its receiving part. The combination of the aforementioned elements therefore results in an \ac{LPTV} impulse response $h(t,\tau)$ observed at the instant $t$ given the injection of an impulse at $t-\tau$, where $\tau$ is a time lag, of the reflection channel experienced by the \ac{PLM} connected to the network. Assuming that the injection and subsequent capture of reflections of signals takes place within a coherence time, in which variations in loads or any other element of the network are irrelevant, we can consider that the distribution network is an \ac{LTI} system and denote its reflection channel impulse response simply as $h(t)$. We bring the reader's attention to the fact that such assumption is made henceforth and that $h(t)$ is considered to be deterministic in this work, although it could be considered a random process under certain circumstances.

The impulse response of the reflection channel $h(t)$, as its name suggests, is intimately related  to reflections suffered by a signal propagating through the distribution grid, which are raised by impedance discontinuities along the power supply cable. Such reflections are quantified via the input reflection coefficient between the \ac{PLM} of output impedance $Z_{PLC}(f)$ and the distribution grid of input impedance $Z_{in}(f)$, which is expressed as \cite{paul2007}
\begin{equation}\label{eq:Gamma_in}
\Gamma_{in}(f) = \frac{Z_{in}(f)-Z_{PLC}(f)}{Z_{in}(f)+Z_{PLC}(f)}.
\end{equation}
From $\Gamma_{in}(f)$, we finally obtain the impulse response of the reflection channel via the inverse Fourier transform, i.e.,
\begin{equation}\label{eq:h(t)}
h(t) = \int_{-\infty}^{\infty}\Gamma_{in}(f)e^{j2\pi ft}df.
\end{equation}

If no abnormality occurs in the distribution network, i.e., its topology is just as depicted in Fig~\ref{fig:dist-network}, we say that the network is under normal operation conditions, presenting input impedance $Z_{in,normal}(f)$ and consequently input reflection coefficient $\Gamma_{in,normal}(f)$ and impulse response of the reflection channel $h_{normal}(t)$.
However, if a localized fault with impedance $Z_{fault}(f)$ occurs, then the topology is changed to the one depicted in Fig~\ref{fig:dist-network-fault}. As a consequence, the input impedance of the distribution network changes to $Z_{in,fault}(f)$ and additional reflections are suffered by the signal propagating in this network, resulting in an input reflection coefficient $\Gamma_{in,fault}(f)$ and an impulse response $h_{fault}(t)$ of the reflection channel. In addition to the normal reflections caused by impedance mismatchings and load connections already present in $h_{normal}(t)$, the impulse response $h_{fault}(t)$ also encompasses reflections raised by the occurrence of the fault. 

The detection and location of the fault can therefore be performed via an analysis of the reflections raised by the impedance discontinuity caused by the fault. For this purpose, a \ac{TDR} system operating over the distribution network is used for obtaining a reflectogram $\rho(t)$, which is an estimate of $h(t)$. Our \ac{TDR} system operates on the system depicted in Figure~\ref{fig:h_circulator}. The transmitting part of the \ac{PLC} transceiver injects a pulse $x(t)$ into the distribution network with reflection channel impulse response $h(t)$. This results in the reflected signal $\tilde{y}(t)=x(t)\star h(t)$, where $\star$ is the convolution operator. The captured signal by the receiving part of the transceiver is composed by $\tilde{y}(t)$ plus an additive noise $v(t)$, which is a zero-mean \ac{WSS} random process, and is represented by $y(t)=\tilde{y}(t)+v(t)$, or, alternatively,
\begin{equation}\label{eq:conv_noise}
y(t) = x(t)\star h(t) + v(t),
\end{equation} 
which is also a \ac{WSS} random process.
If no further processing is performed on $y(t)$, we obtain the reflectogram directly from the received signal, i.e., $\rho(t)=~y(t)$. Note that $\rho(t)$ is also a \ac{WSS} random process.

Under normal and fault conditions, respectively, we obtain the reflectograms $\rho_{normal}(t)$ and $\rho_{fault}(t)$, which are weighted estimates of $h_{normal}(t)$ and $h_{fault}(t)$, respectively. A comparison of these reflectograms allows the identification of emerging impedance discontinuities in the latter reflectogram that were not present in the former.
As $\rho_{normal}(t)$ and $\rho_{fault}(t)$ differ only for the reflections caused by the impedance discontinuity at the fault point, an appropriate parameter for evaluating fault occurrence is defined as
\begin{equation}\label{eq:Delta_t}
\Delta_\rho(t) = \rho_{fault}(t)-\rho_{normal}(t).
\end{equation}
The fault detection problem resumes therefore to setting an amplitude threshold $\gamma$ for $\Delta_\rho(t)$, above which reflections are considered to have been raised by a fault.
For solving the fault location problem, we have
\begin{equation}
\Delta_\rho(d) = \Delta_\rho\left(\frac{v_pt}{2}\right),
\end{equation}
where $d$ is the distance in meters between the measurement point and the impedance discontinuity that caused the reflection, and $v_p$ is the propagation velocity in m/s along the power supply cable. The factor $1/2$ relates the actual distance of the impedance discontinuities from the measurement point to the round trip time of the observed reflections.

As previously mentioned, the occurrence of a fault causes multiple reflections, the first of them being the one that travels in a direct path from the fault point to the measurement point. The further reflections, in their turn, are raised when the former reflected signal passes by successive impedance discontinuities such as at load connection and branch points along its path. Hence, the fault location procedure consists of finding the position associated with the first peak of $\Delta_\rho(d)$ whose amplitude is above the {threshold $\xi$}. This results in an estimate of the distance from the fault to the measurement point, denoted by $d_{fault}$ and expressed by
{\begin{align}
d&_{fault} = \min\{d\}\nonumber\\
& \text{subject to}\nonumber\\
& \Delta_\rho(d) > \xi\nonumber\\
& \Delta\rho'(d) = 0,
\end{align}}
where $\Delta_\rho'(d)$ is the derivative of $\Delta_\rho(d)$ with respect to $d$.

Given the described system, the \ac{TDR} procedure aims to obtain a reflectogram $\rho(t)$ that is an appropriate estimate of $h(t)$. In order for this to happen, we would ideally have $x(t)=\delta(t)$, where $\delta(t)$ is the Dirac delta function, and $v(t)=0$. The corresponding continuous frequency-domain representation for $x(t)$ would therefore be, i.e., $X(f) = 1$, which denotes a zero-phase signal with constant magnitude in the continuous frequency-domain. In spite of such $x(t)$ resulting in the best achievable estimate of $h(t)$, transmitting such an ideal signal is unfeasible, as a practical \ac{TDR} system is band-limited. 
Considering that our \ac{TDR} system operates in the frequency band from $0$ to $B$ Herz, Section~\ref{sec:pc_tdr} describes a band-limited \ac{TDR} system that allows obtaining an appropriate estimate of $h(t)$ via the transmission of pulses that can be \textit{a priori} designed to comply with the aforementioned constraints.

\section{Pulse-Compression TDR}\label{sec:pc_tdr}

When obtaining a reflectogram, it is desired to have a transmit pulse $x(t)$ that is sufficiently narrow to provide a fair approximation of $\delta(t)$ and therefore adequate temporal and consequently spatial resolution for the reflectogram $\rho(t)$. It is also desired to have a high transmission power, so that the receiving part of the reflectometer experiences an adequate \ac{SNR}. 

A first approach to meet those requirements in a band-limited system is to send a narrow pulse, with well-defined main lobe and sidelobes. As we aim to have $x(t)$ that approximates $\delta(t)$, the side lobes should present significantly lower amplitude than the main lobe to avoid undesirable distortions in the reflectogram. Although this would allow practical implementation, we would have a transmit pulse $x(t)$ with high \ac{PAPR}, i.e., great part of its energy would be concentrated in a short time interval corresponding to the main lobe of the transmit pulse. This may be undesired, since there are upper bounds on the maximum amplitude of the transmit pulse due to hardware limitations and regulatory constraints.

A solution to this issue is transmitting a longer $x(t)$ that can present lower \ac{PAPR}, i.e., does not need to present well-defined main lobe and sidelobes, and then performing processing on the received signal $y(t)$ to obtain the reflectogram $\rho(t)$. This processing, called pulse compression \cite{chang2015,richardsvol1,temes1962}, artificially turns the originally transmit pulse $x(t)$ into a narrow pulse with high \ac{PAPR} that in its turn approximates $\delta(t)$, improving the resolution and elevating the \ac{SNR} of the reflectogram $\rho(t)$.
As the pulse-compression \ac{TDR} procedure, i.e., the obtaining of reflectograms, is the same under both normal and fault operating conditions, we henceforth omit the subscripts from the reflection channel impulse response and the reflectogram and denote them simply as $h(t)$ and $\rho(t)$.

\begin{figure}[!b]
	\centering
	\psfrag{r}[c][c][.85]{$p(t)$}
	\psfrag{R}[c][c][.85]{$R_{pp}(\tau)$}
	\psfrag{c}[c][c][.85]{$p^*(-t)$}
	\includegraphics[width=6.5cm]{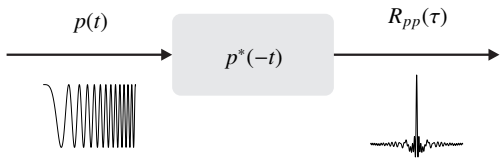}
	\caption{Pulse compression procedure.} \label{fig:pulse_compression}
\end{figure}

For describing the pulse compression procedure, let the transmit pulse $x(t)$ be expressed as the normalization of a pulse $p(t)$ by its 2-norm, followed by a multiplication by an energy factor, i.e.,
\begin{equation}\label{eq:s_p}
x(t) = \sqrt{E}\frac{p(t)}{\|p(t)\|},
\end{equation}
where $E$ is the energy of $x(t)$ and $\|\cdot\|$ returns the 2-norm of a signal. Therefore, the pulse compression procedure consists of performing the convolution of the received signal $\rho(t)$ with a pulse compression filter whose impulse response is $p^*(-t)/\|p^*(-t)\|$, where $(\cdot)^*$ is the complex conjugate operator. The result of such operation is the reflectogram, which can be expressed as
\begin{eqnarray}
\rho(t) & = & y(t)\star \frac{p^*(-t)}{\|p^*(-t)\|} \nonumber\\
& = & \left(\sqrt{E}\frac{p(t)}{\|p(t)\|}\star h(t) \right)\star \frac{p^*(-t)}{\|p^*(-t)\|}  + \nonumber\\
&  & v(t)\star \frac{p^*(-t)}{\|p^*(-t)\|}.
\end{eqnarray}
Knowing that
\begin{eqnarray}
R_{pp}(\tau) & = & \int_{-\infty}^{\infty}p(t)p^*(t-\tau)dt\nonumber\\
& = & p(t)\star p^*(-t)|_{t=\tau}
\end{eqnarray}
is the autocorrelation function of the deterministic energy signal $p(t)$ at the time lag $\tau$ \cite{richardsvol1}, we can write
\begin{equation}\label{eq:system-autocorr}
\rho(t) = \frac{\sqrt{E}}{\|p(t)\|^2}R_{pp}(\tau)\star h(t) + v(t)\star \frac{p^*(-t)}{\|p^*(-t)\|}.
\end{equation}
As the autocorrelation of a signal is pulse-shaped, one can easily see from \eqref{eq:system-autocorr} that the pulse compression filter is used with the purpose of compressing $p(t)$ into the pulse $R_{pp}(\tau)$, which is depicted in Figure~\ref{fig:pulse_compression}. As $R_{pp}(\tau)=~p(t)\star p^*(-t)|_{t=\tau}$, its resulting continuous frequency-domain representation is \mbox{$R_{pp}(f)=P(f)P^*(f)$}, where $P(f)$ is the continuous frequency-domain representation of $p(t)$. Alternatively,
\begin{equation}
R_{pp}(f)=\left|P(f)\right|^2,
\end{equation}
which is a zero-phase signal and can be designed in order to approximate the ideal scenario of a flat continuous frequency-domain representation described in Section~\ref{sec:statement}.

If $R_{pp}(\tau)$ has a sufficiently narrow main lobe and sidelobes with sufficiently low amplitude, then it can be assumed that
\begin{equation}\label{eq:channel_est}
\rho(t) \approx \frac{\sqrt{E}}{\|p(t)\|^2}h(t) + v(t)\star \frac{p^*(-t)}{\|p^*(-t)\|}.
\end{equation}
We can see that the first term of the sum from \eqref{eq:channel_est} is a weighted version of $h(t)$, while the second term refers to the noise component. Despite having its amplitude altered over time, the noise component will have the same power, as it is convolved with the unit-power compression filter.

Based on the carried-out discussion, one can easily conclude that the quality of a reflectogram will depend upon two main factors, namely \textit{(i)} how well $R_{pp}(\tau)$ approximates $\delta(t)$, i.e. how adequate the estimate of $h(t)$ is, and \textit{(ii)} the \ac{SNR} level, which drives the performance of the threshold detector previously described. The former depends only on the transmit pulse's characteristics, while the latter depends both on the effectiveness of the pulse compression procedure and on the additive noise characteristics. In this work, we focus on aspects of the reflectogram that do not depend on specific system's conditions, being therefore the sole investigation of autocorrelation functions of typical \ac{PLC} pulses our subject of study. In this sense, expressions for $R_{pp}(\tau)$ of such signals are derived in Section~\ref{sec:PLC} in order to enable an assessment of their suitability for \ac{TDR} purposes.

\section{Compression of PLC Pulses}\label{sec:PLC}

In this section, we present pulses of four commonly adopted modulation techniques for \ac{PLC} systems and derive their autocorrelation functions in order to determine whether they are suitable for \ac{TDR} purposes.

\subsection{Hermitian-Symmetric OFDM}\label{subsec:ofdm}

Let an \ac{HS-OFDM} system transmit a complex symbol \mbox{$\mathbf{P}=[P_0, P_1, \cdots, P_{N-1}]$}, where $P_k\in\mathbb{C}$, $k=0,\cdots,N-1$. By performing an Hermitian symmetric mapping \cite{giroto2018} on $\mathbf{P}$ before digital-to-analog conversion, the transmit pulse in the continuous time-domain can be expressed as
\begin{eqnarray}\label{eq:hs_ofdm}
p_{OFDM}(t) & = & \Re\{P_{N-1}\}e^{j2\pi 0\Delta f t} \nonumber\\
& + & \sum_{k=0}^{N-2}\left(P_ke^{j2\pi k\Delta f t}+P_k^*e^{-j2\pi k\Delta f t}\right) \nonumber\\
& + & \Im\{P_{N-1}\}e^{j2\pi N\Delta f t},
\end{eqnarray}
where $-T/2<t<T/2$, with $T$ denoting the pulse duration, $\Delta f=1/T$ is the subcarrier frequency spacing, and $\Re\{\cdot\}$ and $\Im\{\cdot\}$ respectively denote the real and imaginary parts of a complex number.  An analysis of \eqref{eq:hs_ofdm} reveals that $p(t)$ occupies a frequency bandwidth equal to
\begin{equation}
B_{OFDM} = N\Delta f,
\end{equation}
i.e., it ranges from $-B_{OFDM}$ to $B_{OFDM}$ in the continuous frequency-domain.

Assuming uniform power allocation and \ac{BPSK} modulation, we have $P_k\in\mathbb{R}, \forall n$. This results in $\left|P_k\right|=\sqrt{E}$, as well as in $P_k=P_k^*$ and $\Im\{P_{N-1}\}=0$. Therefore, the autocorrelation function of $p(t)$ for an \ac{HS-OFDM} system is defined for $-T<t<T$ as in \eqref{eq:Rpp_ofdm}.

\begin{figure*}[!t]
	\small 
	\begin{eqnarray}\label{eq:Rpp_ofdm}
	R_{pp,OFDM}(\tau) & \triangleq & P_{N-1}^2(2T-|\tau|) + P_{N-1}\sum_{l=0}^{N-2}P_l\left(\frac{\sin(2\pi l\Delta fT)+\sin[2\pi l\Delta f(T-|\tau|)]+\sin[2\pi l\Delta f(2T-|\tau|)]}{\pi\Delta f l}\right)\\\nonumber
	& + & \sum_{l=0}^{N-2}\sum_{m=0}^{N-2}P_lP_m\left(\frac{\sin[2\pi l\Delta f(T-|\tau|)]+\sin[2\pi m\Delta f(T-|\tau|)]}{\pi\Delta f(l+m)}+\frac{\sin[2\pi l\Delta f(T-|\tau|)]-\sin[2\pi m\Delta f(T-|\tau|)]}{\pi\Delta f(l-m)}\right)
	\end{eqnarray}
	\vspace*{4pt}
	\hrulefill
	\vspace*{4pt}
\end{figure*}

\begin{figure}[!t]
	\centering
	\subfloat[ ]{
		\psfrag{XX}[c][c]{$t$}
		\psfrag{YY}[c][c]{$\overline{p}_(t)$}
		\includegraphics[width=4.25cm]{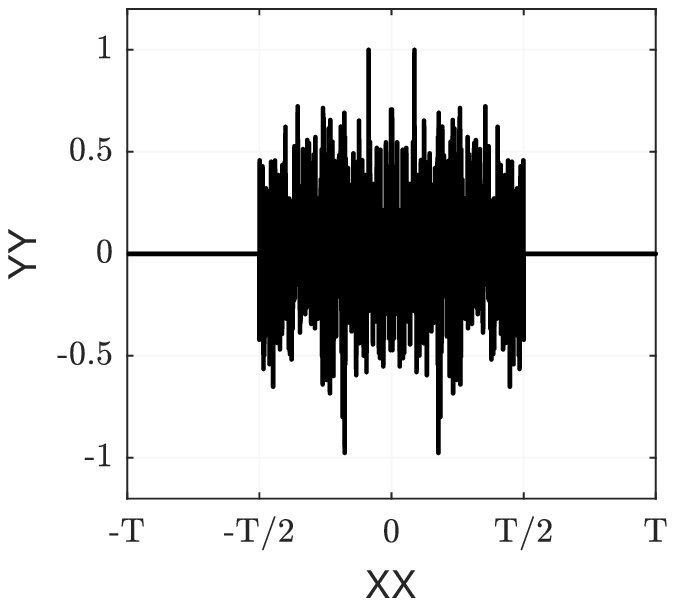}
		\label{fig:p1}
	}	
	\subfloat[ ]{
		\psfrag{XX}[c][c]{$\tau$}
		\psfrag{YY}[c][c]{$\overline{R}_{p,OFDM}(\tau)$}
		\includegraphics[width=4.25cm]{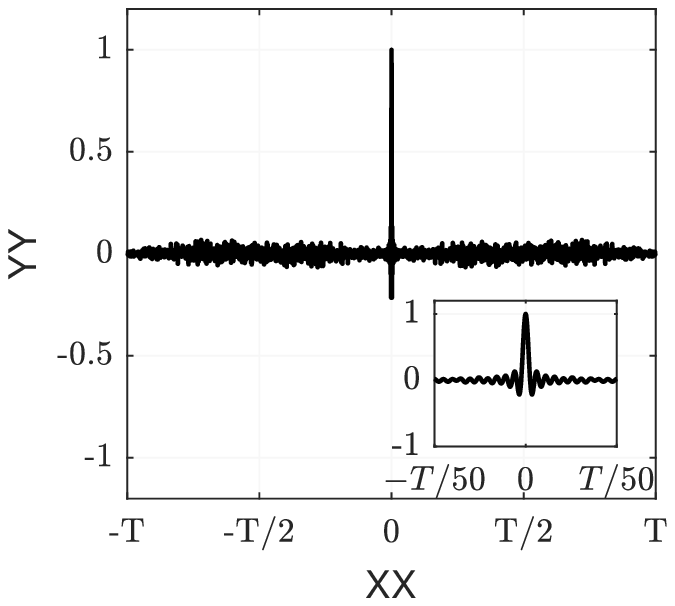}
		\label{fig:Rp1}
	}
	\caption{HS-OFDM pulse: (a) $\overline{p}_{OFDM}(t)$ and (b) its autocorrelation $\overline{R}_{p,OFDM}(\tau)$.}\label{fig:pulse1}
\end{figure}

Figs.~\ref{fig:p1} and \ref{fig:Rp1} show respectively the normalized pulse and its normalized autocorrelation for the \ac{HS-OFDM} scheme. In this figure, the normalized \ac{HS-OFDM} pulse and its autocorrelation function are respectively given by $\overline{p}_{OFDM}(t)=p_{OFDM}(t)/\max\{\left|p_{OFDM}(t)\right|\}$ and $\overline{R}_{p,OFDM}(\tau)~=~R_{pp,OFDM}(\tau)/\max\{\left|R_{pp,OFDM}(\tau)\right|\}$, where $\max\{\cdot\}$ returns the maximum value of a signal. This normalization will also be performed for presenting the pulses and their autocorrelations for the remaining cases.

\subsection{Ultra-Wideband}\label{subsec:uwb}

For the \ac{UWB} system, we consider the first \cite{salmento2017} and second \cite{tonello2011} derivatives of the Gaussian pulse with respect to $t$, i.e. $g'(t)$ and $g''(t)$, where $g(t)$ is the Gaussian pulse. The Gaussian pulse is defined for $-T/2<t<T/2$ and is expressed in the continuous time-domain as
\begin{equation}
g(t) = \frac{1}{\sqrt{2\pi}\sigma}e^{-\frac{1}{2}\left(\frac{t}{\sigma}\right)^2},
\end{equation}
with $\sigma$ denoting the variance.

\subsubsection{First derivative of the Gaussian pulse}\label{subsubsec:uwb_1}

The first derivative of the Gaussian pulse is expressed in the continuous time-domain as
\begin{equation}
p_{UWB-1}(t) = -\frac{t}{\sqrt{2\pi}\sigma^3}e^{-\frac{1}{2}\left(\frac{t}{\sigma}\right)^2}.
\end{equation}
Besides $p_{UWB-1}(t)$ being defined for $-T/2<t<T/2$, we have that its effective duration is $T=7\sigma$ \cite{benedetto2006}. The occupied frequency bandwidth by $p_{UWB-1}(t)$ is defined as the lowest frequency beyond which its continuous frequency-domain representation $P_{UWB-1}(f)$ is at least $30$~dB below its maximum value \cite{tonello2011}. Based on this assumption, we have a continuous frequency-domain representation ranging from $-B_{UWB-1}$ to $B_{UWB-1}$, in which
\begin{equation}
B_{UWB-1} \approx \frac{1599.49\times 10^{-3}}{\pi\sigma}.
\end{equation}

The autocorrelation function of $p_{UWB-1}(t)$ for the first derivative of the Gaussian pulse is therefore given by
\begin{equation}\label{eq:Rpp_uwb_1}
R_{pp,UWB-1}(\tau) = \frac{-1}{8\sqrt{\pi}\sigma^5}\left(\tau^2-2\sigma^2\right)e^{-\frac{\tau^2}{4\sigma^2}}
\end{equation}
and is defined for $-T<\tau<T$, or, alternatively, $-7\sigma<\tau<7\sigma$.

Figs.~\ref{fig:p2} and \ref{fig:Rp2} show respectively the normalized pulse $\overline{p}_{UWB-1}(t)$ and its normalized autocorrelation $\overline{R}_{p,UWB-1}(\tau)$ for the first derivative of the Gaussian pulse.

\subsubsection{Second derivative of the Gaussian pulse}\label{subsubsec:uwb_2}

The second derivative of the Gaussian pulse is expressed in the continuous time-domain as
\begin{equation}
p_{UWB-2}(t) = \frac{t^2-\sigma^2}{\sqrt{2\pi}\sigma^5}e^{-\frac{1}{2}\left(\frac{t}{\sigma}\right)^2}.
\end{equation}
and is defined for $-T/2<t<T/2$, with $T=7\sigma$ as in the case of $p_{UWB-1}(t)$. Adopting the same criterion from the previous \ac{UWB} pulse, the occupied frequency bandwidth by $p_{UWB-2}(t)$ is given by
\begin{equation}
B_{UWB-2} \approx \frac{564.65\times 10^{-3}}{\sigma},
\end{equation}
i.e. it continuous frequency-domain representation ranges from $-B_{UWB-2}$ to $B_{UWB-2}$.

By convolving $p_{UWB-2}(t)$ with $p_{UWB-2}^*(-t)$, one obtains the autocorrelation function of $p_{UWB-2}(t)$ for the second derivative of the Gaussian pulse, expressed as
\begin{equation}\label{eq:Rpp_uwb_2}
R_{pp,UWB-2}(\tau) = \frac{1}{32\sqrt{\pi}\sigma^9}\left(\tau^4-12\sigma^2\tau^2+12\sigma^4\right)e^{-\frac{\tau^2}{4\sigma^2}},
\end{equation}
which is defined for $-T<\tau<T$, or, alternatively, $-7\sigma<\tau<7\sigma$.

Figs.~\ref{fig:p3} and \ref{fig:Rp3} show respectively the normalized pulse $\overline{p}_{UWB-2}(t)$ and its normalized autocorrelation $\overline{R}_{p,UWB-2}(\tau)$ for the second derivative of the Gaussian pulse.

\subsection{Chirp Spread Spectrum}\label{subsec:chirp}

The baseband \ac{CSS} system multiplies an up- or down-chirp $p_{CSS}(t)$, properly normalized to present unit power, by a factor $\sqrt{E}$ before transmitting it. In this work, we consider the baseband up-chirp defined for $-T/2<t<T/2$ and denoted in the continuous time-domain by
\begin{equation}
p_{CSS}(t) = \cos(\pi\mu t^2).
\end{equation}
where $\mu\in\mathbb{R}$ is the chirp rate.
For a given $\mu$ value, the occupied frequency bandwidth by $p_{CSS}(t)$ is given by
\begin{equation}
B_{CSS} = |\mu| \frac{T}{2},
\end{equation}
which results in the fact that $p_{CSS}(t)$ linearly sweeps frequencies from $-B_{CSS}$ to $B_{CSS}$ in $T$ seconds.

\begin{figure}[!t]
	\centering
	\subfloat[ ]{
		\psfrag{XX}[c][c]{$t$}
		\psfrag{YY}[c][c]{$\overline{p}_{CSS}(t)$}
		\includegraphics[width=4.2cm]{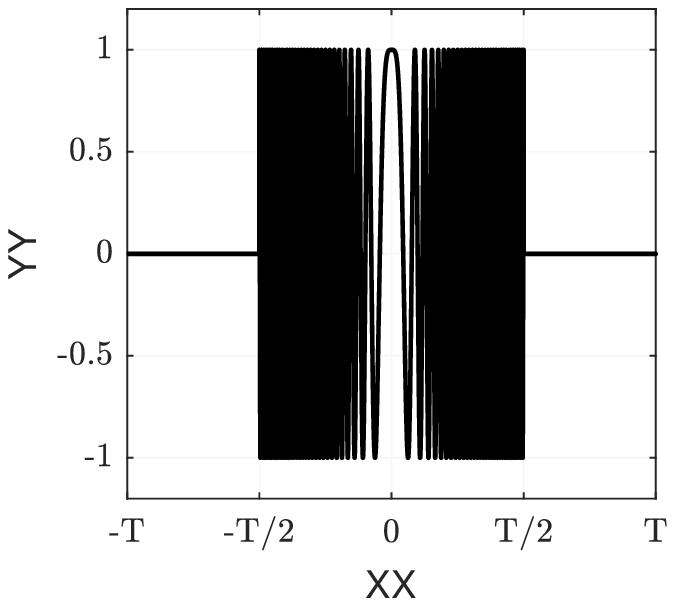}
		\label{fig:p4}
	}	
	\subfloat[ ]{
		\psfrag{XX}[c][c]{$\tau$}
		\psfrag{YY}[c][c]{$\overline{R}_{p,CSS}(\tau)$}
		\includegraphics[width=4.2cm]{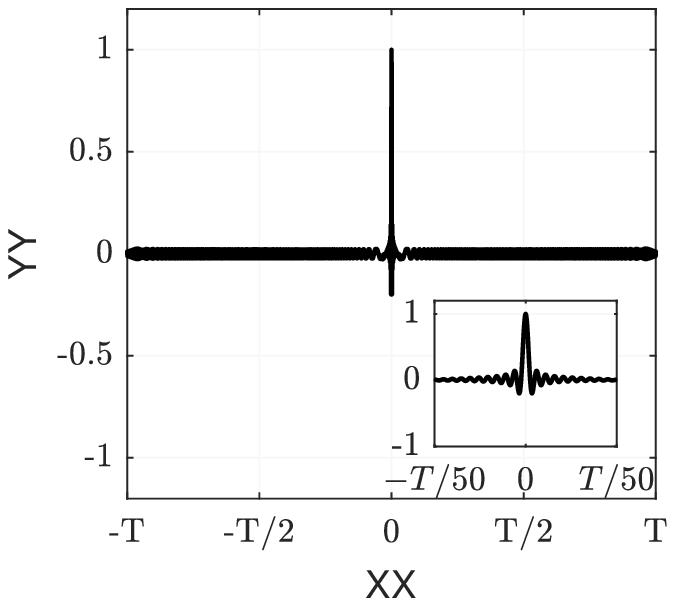}
		\label{fig:Rp4}
	}
	\caption{\ac{CSS} pulse: (a) $\overline{p}_{CSS}(t)$ and (b) its autocorrelation $\overline{R}_{p,CSS}(\tau)$.}\label{fig:pulse4}
\end{figure}

By convolving $p_{CSS}(t)$ with its matched counterpart $p_{CSS}^*(-t)$, one obtains $R_{pp,CSS}(\tau)$, which is defined for $-T~<~\tau<T$ and is defined as
\begin{eqnarray}\label{eq:Rpp_chirp}
R_{pp,CSS}(\tau) & \triangleq & \Lambda(T/2,\tau) - \Lambda(-T/2,\tau)\nonumber\\
& + & \frac{1}{2\pi\mu \tau} \cos(\pi\mu \tau^2) \sin(\pi\mu \tau T),
\end{eqnarray}
with
\begin{eqnarray}
\Lambda(t,\tau) & \triangleq & \frac{1}{4\sqrt{\mu}} \left[\cos\left(\frac{\pi\mu \tau^2}{2}\right) \mathcal{C}\left(\sqrt{\mu}(2t-\tau)\right) \right.\nonumber\\
& - & \left. \sin\left(\frac{\pi\mu \tau^2}{2}\right) \mathcal{S}\left(\sqrt{\mu}(2t-\tau)\right)\right],
\end{eqnarray}
where $\mathcal{C}\left(\cdot\right)$ and $\mathcal{S}\left(\cdot\right)$ denote, respectively, the Fresnel cosine and Fresnel sine integrals as in their second definition in \cite{gradshteyn1980}.
Figs.~\ref{fig:p4} and \ref{fig:Rp4} show respectively the normalized pulse $\overline{p}_{CSS}(t)$ and its normalized autocorrelation $\overline{R}_{p,CSS}(\tau)$ for the considered up-chirp pulse.

\begin{figure}[!t]
	\centering
	\subfloat[ ]{
		\psfrag{XX}[c][c]{$t$}
		\psfrag{YY}[c][c]{$\overline{p}_{UWB-1}(t)$}
		\includegraphics[width=4.25cm]{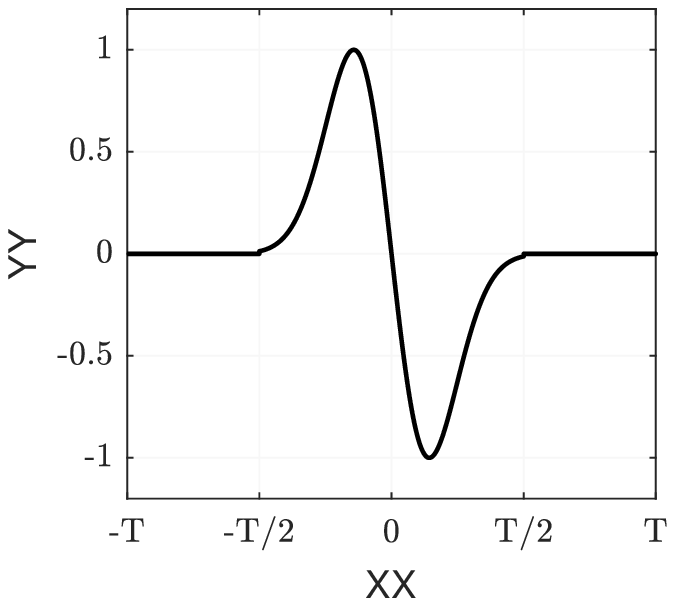}
		\label{fig:p2}
	}	
	\subfloat[ ]{
		\psfrag{XX}[c][c]{$\tau$}
		\psfrag{YY}[c][c]{$\overline{R}_{p,UWB-1}(\tau)$}
		\includegraphics[width=4.25cm]{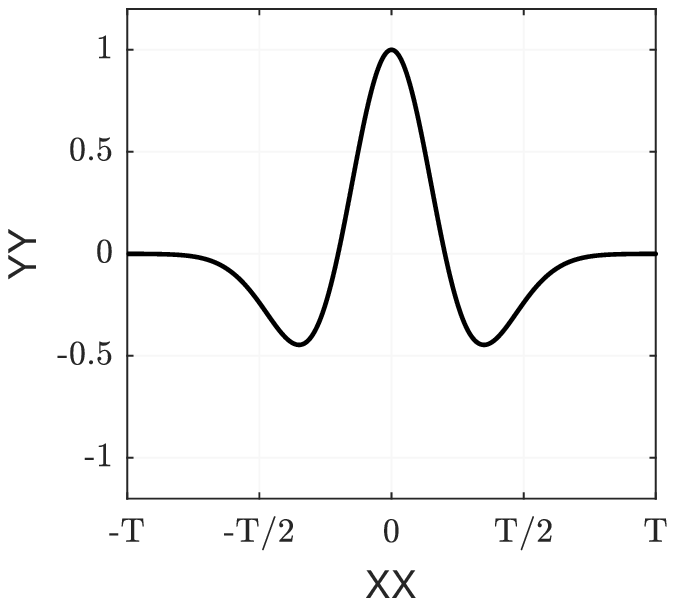}
		\label{fig:Rp2}
	}
	\caption{First derivative of the Gaussian pulse: (a) $\overline{p}_{UWB-1}(t)$ and (b) its autocorrelation $\overline{R}_{p,UWB-1}(\tau)$.} \label{fig:pulse2}
\end{figure}

\section{Reflectogram Quality}\label{sec:quality}

Disregarding noise effect, the quality of the estimate of $h(t)$ provided by the reflectogram $y(t)$ according to \eqref{eq:channel_est} can be assessed by means of different metrics. These metrics can be used for evaluating the range resolution of the reflectometry system, which is the capability of resolving subsequent reflections, as well as the pulse compression effectiveness and the reflectogram distortion caused by the sidelobes of the transmit pulse's autocorrelation function.

\subsection{Range Resolution}\label{subsec:resolution}

\begin{figure}[!t]
	\centering
	\subfloat[ ]{
		\psfrag{A}[c][c][.85]{$-\Tau_{\delta}$}
		\psfrag{B}[c][c][.85]{$0$}
		\psfrag{C}[c][c][.85]{$\Tau_{\delta}$}
		\psfrag{D}[c][c][.85]{$t$}
		\includegraphics[width=6.5cm]{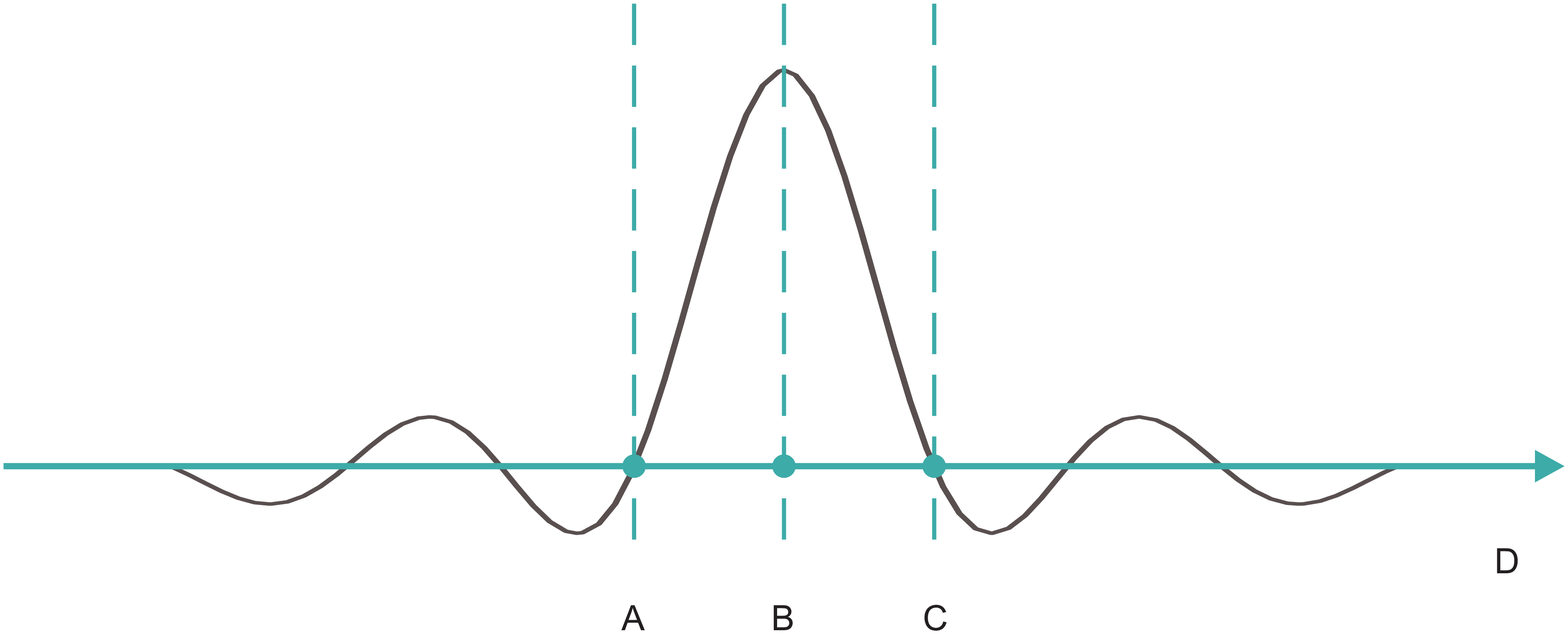}\label{fig:autocorr}
	}\\
	
	\subfloat[ ]{
		\psfrag{D}[c][c][.85]{$t$}
		\psfrag{T}[c][c][.85]{$\Tau_{\delta}$}
		\includegraphics[width=6.5cm]{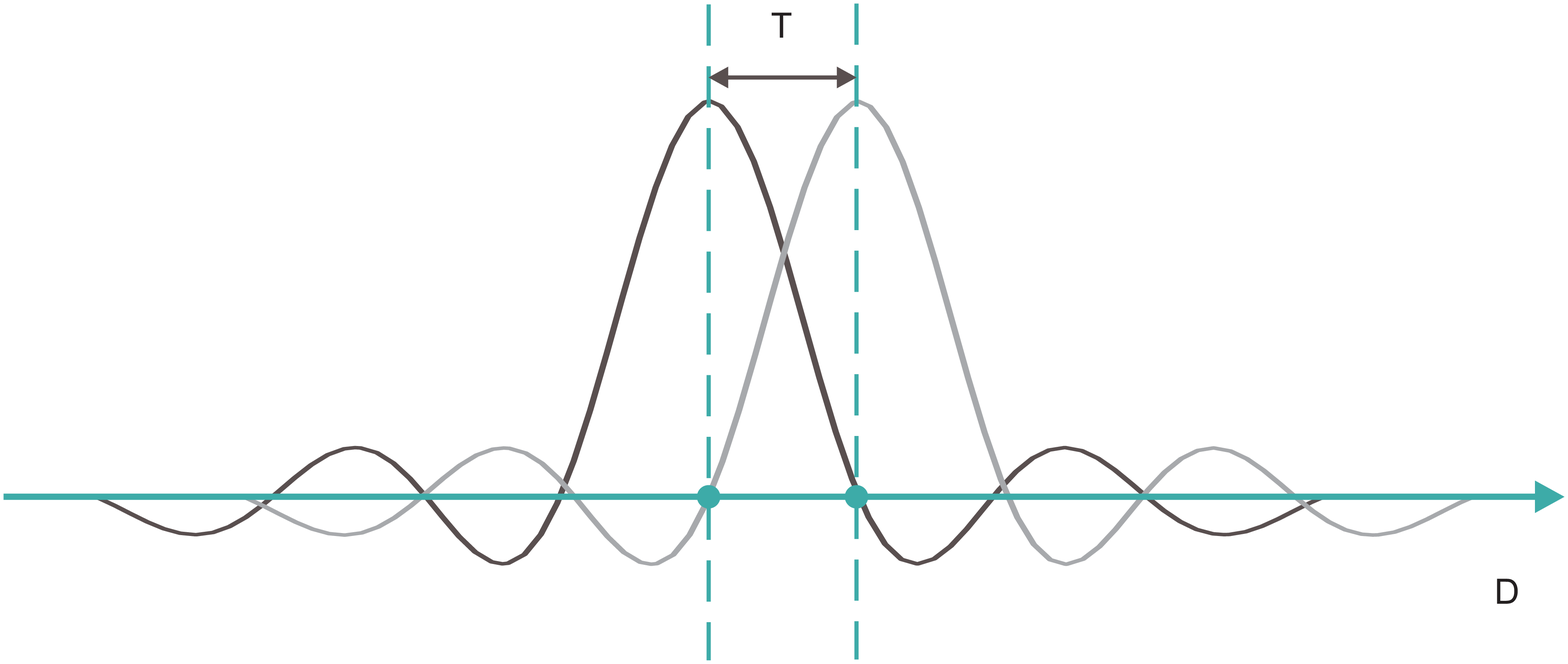}\label{fig:rayleigh_res}	
	}\\
	\caption{Autocorrelation function $R_{pp}(\tau)$: (a) zero crossings and (b) Rayleigh resolution.} 
\end{figure}

In this work, the metric used for evaluating the range resolution is the Rayleigh criterion, widely used in radar systems \cite{richardsvol1}. Given a pulse with main lobe width of $2\Tau_{\delta}$ as shown in Figure~\ref{fig:autocorr}, this criterion states that reflections related to two impedance discontinuities with spacing $\delta$ from one another are only resolved if the peak of the main lobe of one reflections falls at the same time instant as a zero crossing immediately next to the main lobe of the other reflection, i.e. the reflections are spaced from $\Tau_{\delta}=\delta/(v_p/2)$, as depicted in Figure~\ref{fig:rayleigh_res}. This criterion can be used as a metric for comparing the resolution of reflectograms obtained via transmit pulses with different autocorrelation functions.

\begin{figure}[!t]
	\centering
	\subfloat[ ]{
		\psfrag{XX}[c][c]{$t$}
		\psfrag{YY}[c][c]{$\overline{p}_{UWB-2}(t)$}
		\includegraphics[width=4.25cm]{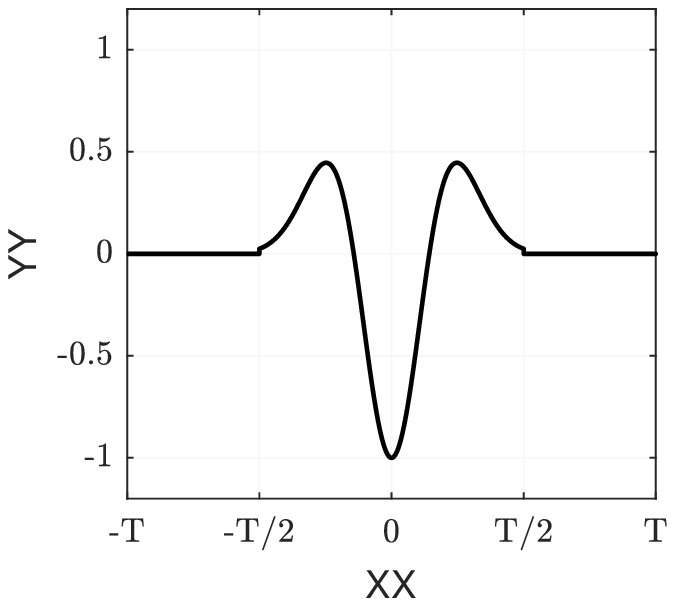}
		\label{fig:p3}
	}	
	\subfloat[ ]{
		\psfrag{XX}[c][c]{$\tau$}
		\psfrag{YY}[c][c]{$\overline{R}_{p,UWB-2}(\tau)$}
		\includegraphics[width=4.25cm]{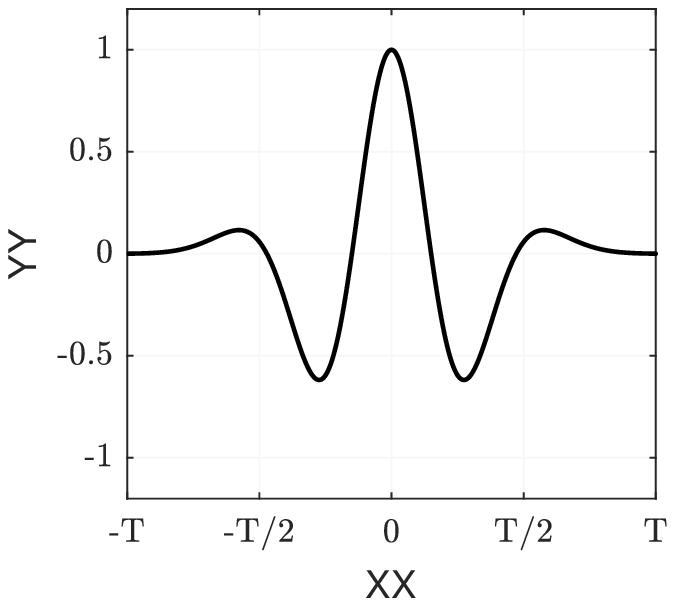}
		\label{fig:Rp3}
	}
	\caption{Second derivative of the Gaussian pulse: (a) $\overline{p}_{UWB-2}(t)$ and (b) its autocorrelation $\overline{R}_{p,UWB-2}(\tau)$.}\label{fig:pulse3}
\end{figure}

The range resolutions associated to the autocorrelation $R_{pp,i}(\tau)$, $i~\in~\{OFDM,UWB-1,UWB-2,CSS\}$, is therefore the product of the time instant of the first positive zero crossing of $R_{pp,i}(\tau)$ by $v_p/2$, i.e. $\delta_i=v_p\Tau_{\delta,i}/2$. Finding the first zero crossings of $R_{pp,i}(\tau)$ resumes to solving $R_{pp,i}(\tau) = 0$ for $\tau$, setting $\Tau_{\delta,i}$ as the smallest absolute value of $\tau$ obtained, and multiplying it by $v_p/2$. The range resolutions for \ac{HS-OFDM}, \ac{UWB} and \ac{CSS} pulses are presented as follows:

\subsubsection{Range Resolution of HS-OFDM Pulse}\label{subsubsec:res_OFDM}

For the \ac{HS-OFDM} pulse with uniform power allocation, $\Tau_{\delta,OFDM}$ is obtained by equaling \eqref{eq:Rpp_ofdm} to zero, yielding $\Tau_{\delta,OFDM}~=~1/(2B_{OFDM})$ \cite{hassen2015,lellouch2016}. Multiplying the result of this procedure by $v_p/2$, we obtain the range resolution $\delta_{OFDM}=v_p\Tau_{\delta,OFDM}/2$, expressed as
\begin{equation}
\delta_{OFDM} = \frac{v_p}{4B_{OFDM}}.
\end{equation}

\subsubsection{Range Resolution of Ultra-Wideband Pulses}\label{subsubsec:res_UWB}

Regarding \ac{UWB} pulses, the Rayleigh resolution $\delta_{UWB-1}=v_p\Tau_{\delta,UWB-1}/2$ for the first derivative of the Gaussian pulse is obtained by equaling \eqref{eq:Rpp_uwb_1} to zero and solving for $\tau$, which yields $\Tau_{\delta,UWB-1}=\sqrt{2}\sigma$. Multiplying this result by $v_p/2$, we get
\begin{eqnarray}
\delta_{UWB-1} = \frac{v_p\sigma}{\sqrt{2}}.
\end{eqnarray}

For the second derivative of the Gaussian pulse, we obtain $\Tau_{\delta,UWB-2}=\sigma\sqrt{6 - 2\sqrt{6}}$
via the same procedure for the previous case. The Rayleigh resolution $\delta_{UWB-2}=v_p\Tau_{\delta,UWB-2}/2$ is then
\begin{equation}
\delta_{UWB-2} = \frac{v_p}{2}\sigma\sqrt{6 - 2\sqrt{6}}.
\end{equation}

\subsubsection{Range Resolution of CSS Pulse}\label{subsubsec:res_CSS}

For the \ac{CSS} case, the expression $R_{pp,CSS}(\tau)=0$ must be solved numerically for $\tau$ and no closed-form expression for $\delta_{CSS}$ could be presented in this case. However, by making use of an approximation for linear chirps with $B_{CSS}T>10$ reported in the literature \cite{richardsvol1}, we obtain $\Tau_{\delta,CSS}\approx1/(2B_{CSS})$ and the range resolution $\delta_{CSS}=v_p\Tau_{\delta,CSS}/2$, which is expressed as
\begin{equation}
\delta_{CSS} \approx \frac{v_p}{4B_{CSS}}.
\end{equation}

\subsection{Pulse compression ratio}\label{subsec:PCR}

Once the range resolution has been obtained, the effectiveness of the pulse compression can be evaluated by means of the \ac{PCR} \cite{eaves1987}, which is a ratio between the pulse length before pulse compression and the effective pulse length after the compression procedure, i.e., $T_{\delta,i}$
\begin{equation}
PCR_i = \frac{T}{T_{\delta,i}},
\end{equation}
with $i~\in~\{OFDM,UWB-1,UWB-2,CSS\}$.

The \ac{PCR} can alternatively be interpreted as a ratio between the effective average power of a pulse compression \ac{TDR} system and a simple pulse system, provided that peak power and range resolution is the same for both cases \cite{eaves1987}. Depending on the noise characteristics, \ac{PCR} can be used for measuring \ac{SNR} gain provided by the pulse compression procedure, being also known as \ac{PCG} \cite{richardsvol1}. Given this context, the \acp{PCR} for \ac{HS-OFDM}, \ac{UWB} and \ac{CSS} pulses are presented as follows:

\subsubsection{Pulse compression ratio of HS-OFDM pulse}\label{subsubsec:PCR_OFDM}

Given the $\Tau_{\delta,OFDM}$ value obtained in Subsection~\ref{subsubsec:res_OFDM}, the \ac{PCR} for the \ac{HS-OFDM} pulse is
\begin{equation}
PCR_{OFDM} = 2B_{OFDM}T.
\end{equation}

\subsubsection{Pulse compression ratio of Ultra-Wideband pulses}\label{subsubsec:PCR_UWB}

For the first derivative of the Gaussian pulse, which presents $\Tau_{\delta,UWB-1}$ as shown in Subsection~\ref{subsubsec:res_UWB}, we have
\begin{eqnarray}
PCR_{UWB-1} & = & \frac{7}{\sqrt{2}} \approx 4.95,
\end{eqnarray}
while the associated \ac{PCR} with the second derivative of the Gaussian pulse is
\begin{eqnarray}
PCR_{UWB-2} & = & \frac{7}{\sqrt{6 - 2\sqrt{6}}} \approx 6.67,
\end{eqnarray}
both results holding for any given $T=7\sigma$.

\subsubsection{Pulse compression ratio of CSS pulse}\label{subsubsec:PCR_chirp}

The $\Tau_{\delta,CSS}$ value obtained in Subsection~\ref{subsubsec:res_CSS} for the \ac{CSS} pulse results in a \ac{PCR} given by
\begin{equation}
PCR_{CSS} \approx 2B_{CSS}T.
\end{equation}

\subsection{Reflectogram distortion}\label{subsec:distortion}

Further metrics can be used for assessing the quality of the reflectogram in terms of distortion caused by the presence of sidelobes in the autocorrelation function of the transmit pulse. In this work, we consider the \ac{PSLR} and \ac{ISLR} metrics \cite{lellouch2016} for this purpose. The \ac{PSLR} measures the ratio
between the highest sidelobe and the main lobe, i.e.,
\begin{equation}\label{eq:pslr}
PSLR = \frac{\max\limits_{\tau\neq 0}\left\{\left|R_{pp}(\tau)\right|\right\}}{\left|R_{pp}(0)\right|}.
\end{equation}
In its turn, the \ac{ISLR} measures the ratio between the integration of the sidelobes and the main lobe, which can be expressed as
\begin{equation}\label{eq:islr}
ISLR = \frac{\int_{-\infty}^{-\Tau_{\delta}}\left|R_{pp}(\tau)\right|d\tau+\int_{\Tau_{\delta}}^{\infty}\left|R_{pp}(\tau)\right|d\tau}{\int_{-\Tau_{\delta}}^{\Tau_{\delta}}\left|R_{pp}(\tau)\right|ad\tau},
\end{equation}
where $\left|\cdot\right|$ is the absolute value operator.

Pulses that present high \ac{PSLR} and/or \ac{ISLR} values can cause significant distortion in the reflectogram if used for \ac{TDR} purposes. However, it is possible to mitigate such undesired effect with signal processing techniques. Among widely used approaches are either windowing the reflectogram, which reduces the level of sidelobes with the drawbacks of \ac{SNR} reduction and broadening of the main lobe \cite{richardsvol1,temes1962,hassen2015}, or designing pulses with desired autocorrelations.

\subsection{Maximum Unambiguous Range}\label{subsec:range}

Besides assessing the quality of the reflectogram in terms of resolution and distortion, it is also important to determine the distance interval within which there is no interference between reflectograms of subsequent pulses. Such interval is named maximum unambiguous range and is related to the maximum duration of the reflection channel $h(t)$, denoted by $T_{h,\max}$, supported by the reflectometry system. 

For this purpose, we define the pulse repetition interval as the time interval between subsequent pulse injections \cite{richardsvol1,eaves1987} down to the distribution network and denote it by $\Delta T_p$. The maximum duration of $h(t)$ is defined based on the fact that the maximum duration of the complete echo response $\rho(t)$, denoted by $T_{\rho,\max}$, is equal to $\Delta T_p$. As $\rho(t)$ is the result of the convolution between $x(t)$ and $h(t)$, we have $T_{\rho,\max}=T+~T_{h,\max}$ and, therefore, $T_{h,\max} = \Delta T_p-T$. As a consequence, the maximum unambiguous range from the measurement point is given by $d_{\max}=v_pT_{h,\max}/2$, i.e.,
\begin{equation}\label{eq:max_range}
d_{\max} = \frac{v_p\left(\Delta T_p-T\right)}{2}.
\end{equation}
It is clear from this expression that the maximum unambiguous range depends upon not only on the duration of the adopted pulse, but also on the periodicity in which it is injected into the channel. An appropriate reflectometry system must therefore consider the duration of a given $h(t)$ in order to define a $T$ and $\Delta T_p$ pair that does not cause interference between reflectograms.

\section{Numerical Analysis}\label{sec:analysis}

In order to investigate the characteristics of the four considered \ac{PLC} pulses for fault sensing via \ac{TDR}, this section assesses their performance in terms of the aspects of range resolution, \ac{PCR}, reflectogram distortion, and maximum unambiguous range. We consider two scenarios for this purpose, namely an European underground low-voltage power distribution network and an US overhead medium-voltage power distribution network in a rural area. For the low-voltage scenario, we consider a power supply cable NAYY150SE with resistance $R'$, inductance $L'$, conductance $G'$, and capacitance $C'$ per unit length calculated as in \cite{lampe_vinck2011}, whereas for the medium-voltage cable, we adopt the power supply cable with distributed parameters listed in \cite{1901_2}.  Based on these parameters, the phase velocity is calculated by $v_p=1/\sqrt{L'C'}$ \cite{paul2007}, resulting in $v_{p,\text{LV}}=1.50\times10^8$ for the considered \ac{LV} cable, and $v_{p,\text{MV}}=2.56\times10^8$ for the considered \ac{MV} cable.

\begin{figure}[!t]
	\vspace{-0.395cm}                       
	\centering
	\subfloat[ ]{
		\psfrag{XX}[c][c]{$T$($\mu$s)}
		\psfrag{YY}[c][c]{$B$(kHz)}
		\psfrag{AAA}[c][c]{\scriptsize 1}
		\psfrag{BBB}[c][c]{\scriptsize 10}
		\psfrag{CCC}[c][c]{\scriptsize 100}
		\psfrag{DDD}[c][c]{\scriptsize 500}
		\psfrag{ZZ}[c][c]{}
		\psfrag{ZZZ}[c][c]{}
		\includegraphics[width=8.5cm]{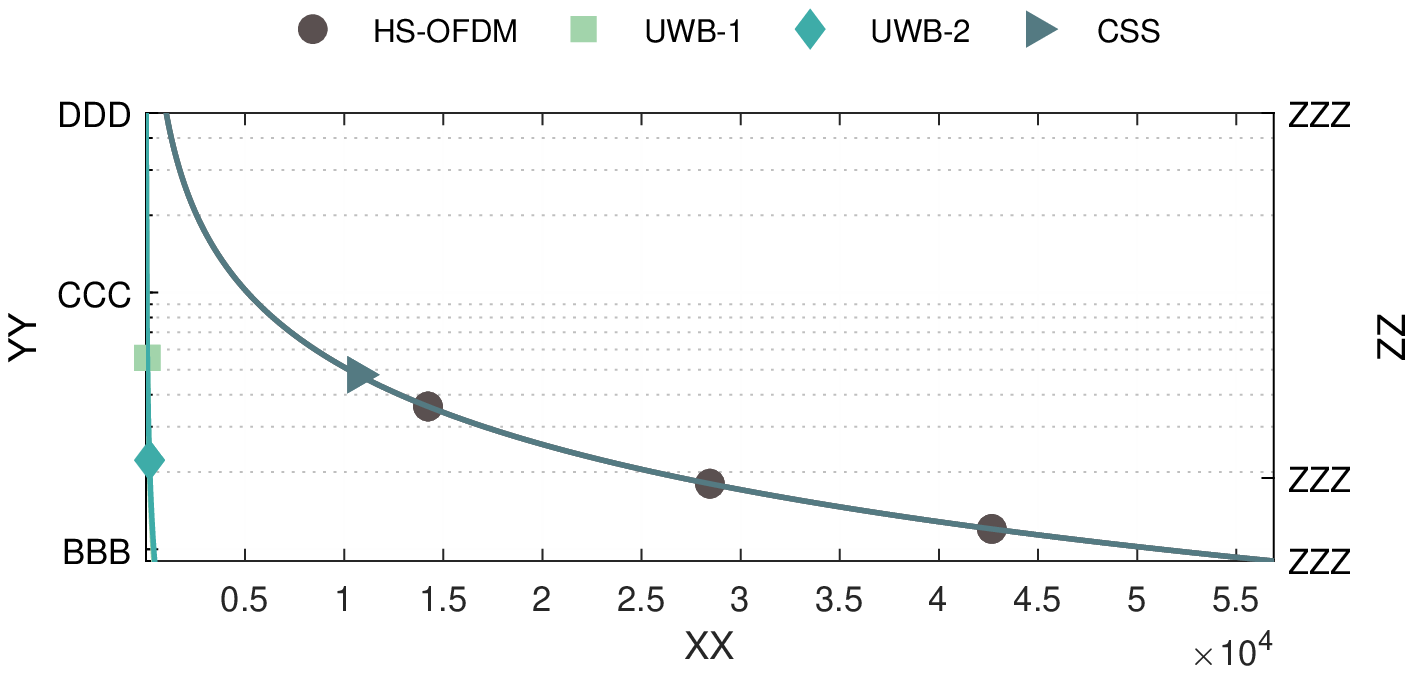}\label{fig:NB_T}
	}\\
	\vspace{-0.395cm}
	\subfloat[ ]{
		\psfrag{XX}[c][c]{$T$($\mu$s)}
		\psfrag{YY}[c][c]{$B$(MHz)}
		\psfrag{AAA}[c][c]{\scriptsize 1.7}
		\psfrag{BBB}[c][c]{\scriptsize 30}
		\psfrag{CCC}[c][c]{\scriptsize 50}
		\psfrag{DDD}[c][c]{\scriptsize 86}
		\psfrag{ZZ}[c][c]{}
		\psfrag{ZZZ}[c][c]{}
		\includegraphics[width=8.5cm]{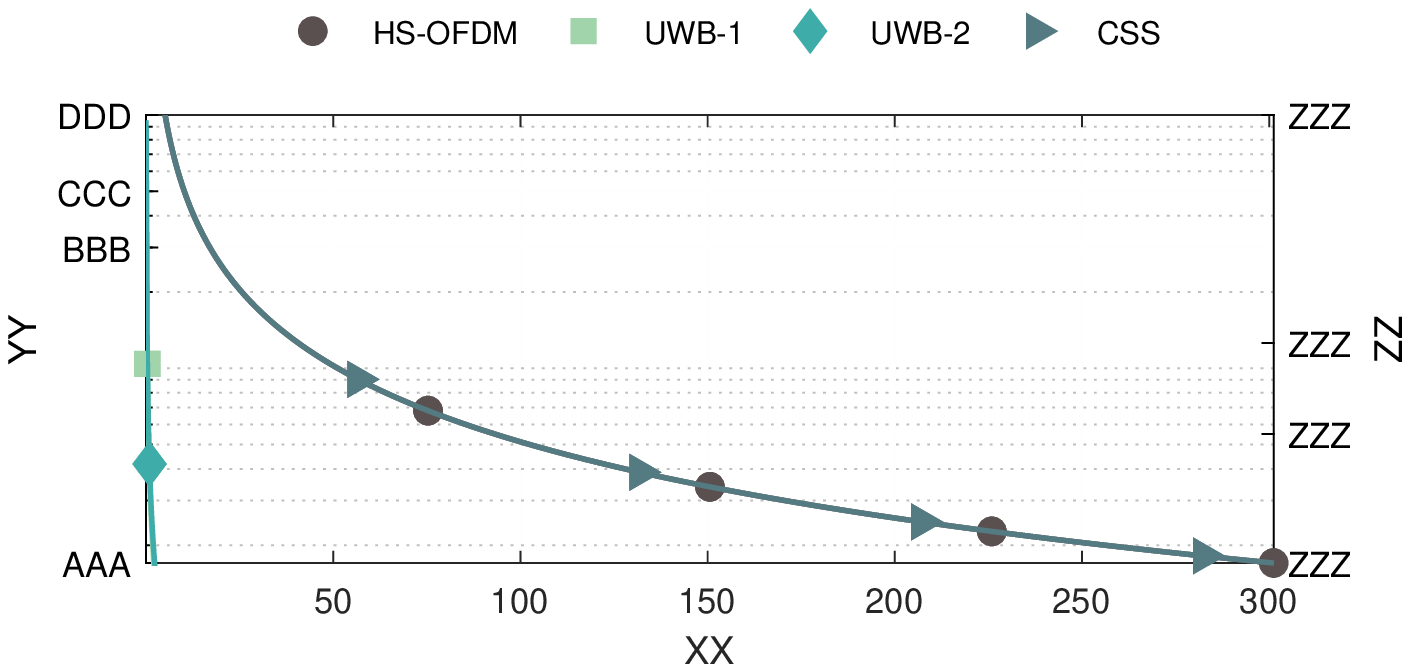}\label{fig:BB_T}
	}\\
	\caption{Occupied frequency bandwidth $B$ as a function of the pulse duration $T$ for (a) \ac{NB}-\ac{PLC} and (b) \ac{BB}-\ac{PLC} systems.}\label{fig:B_T}
\end{figure}

Due to frequency bandwidth occupancy limitations, which belong to \ac{EMC} constraints imposed by regulations, we start our analysis with the influence of the pulse duration $T$ on the occupied frequency bandwidth $B$. Figs.~\ref{fig:NB_T} and \ref{fig:BB_T} show $B$ versus $T$ curves for the four considered \ac{PLC} pulses for \ac{NB}-\ac{PLC} and \ac{BB}-\ac{PLC}, respectively. In these figures, the curves for the \ac{UWB} pulses show very similar behavior with much steeper decay than the curves for \ac{HS-OFDM} and \ac{CSS} pulses, which is due to their pulse-shaped nature. For the carried out analysis in this paper, we consider $N=512$ subcarriers for the \ac{HS-OFDM} scheme and that the \ac{HS-OFDM} and \ac{CSS} pulses have the same duration for a given frequency bandwidth, which implies in increasing $\mu$ along with $B$. This is justified by the fact that \ac{CSS} pulses would present far longer $T$ for high $B$ values if a fixed $\mu$ value is adopted and it results in increasing implementation complexity along with $B$, as larger frequency bandwidths must be swept in shorter time intervals. The observed $\mu$ values, however, are up to $0.98$~kHz/$\mu$s for \ac{NB}-\ac{PLC}, and up to $28.89$~MHz/$\mu$s for \ac{BB}-\ac{PLC} systems, which is considerably smaller than what is reported for applications such as radar \cite{benson2016} and therefore indicates the feasibility of the presented $\mu$ and $B$ pairs.

\begin{figure}[!t]
	\centering
	\subfloat[ ]{
		\psfrag{XX}[c][c]{$T$($\mu$s)}
		\psfrag{YY}[c][c]{$\delta_{\text{LV}}$(km)}
		\psfrag{ZZ}[c][c]{$\delta_{\text{MV}}$(km)}
		\psfrag{AAA}[c][c]{\scriptsize 10$^{\text{-2}}$}
		\psfrag{BBB}[c][c]{\scriptsize 10$^{\text{0}}$}
		\psfrag{CCC}[c][c]{\scriptsize 10$^{\text{2}}$}
		\psfrag{DDD}[c][c]{\scriptsize 10$^{\text{-2}}$}
		\psfrag{EEE}[c][c]{\scriptsize 10$^{\text{0}}$}
		\psfrag{FFF}[c][c]{\scriptsize 10$^{\text{2}}$}
		\includegraphics[width=8.5cm]{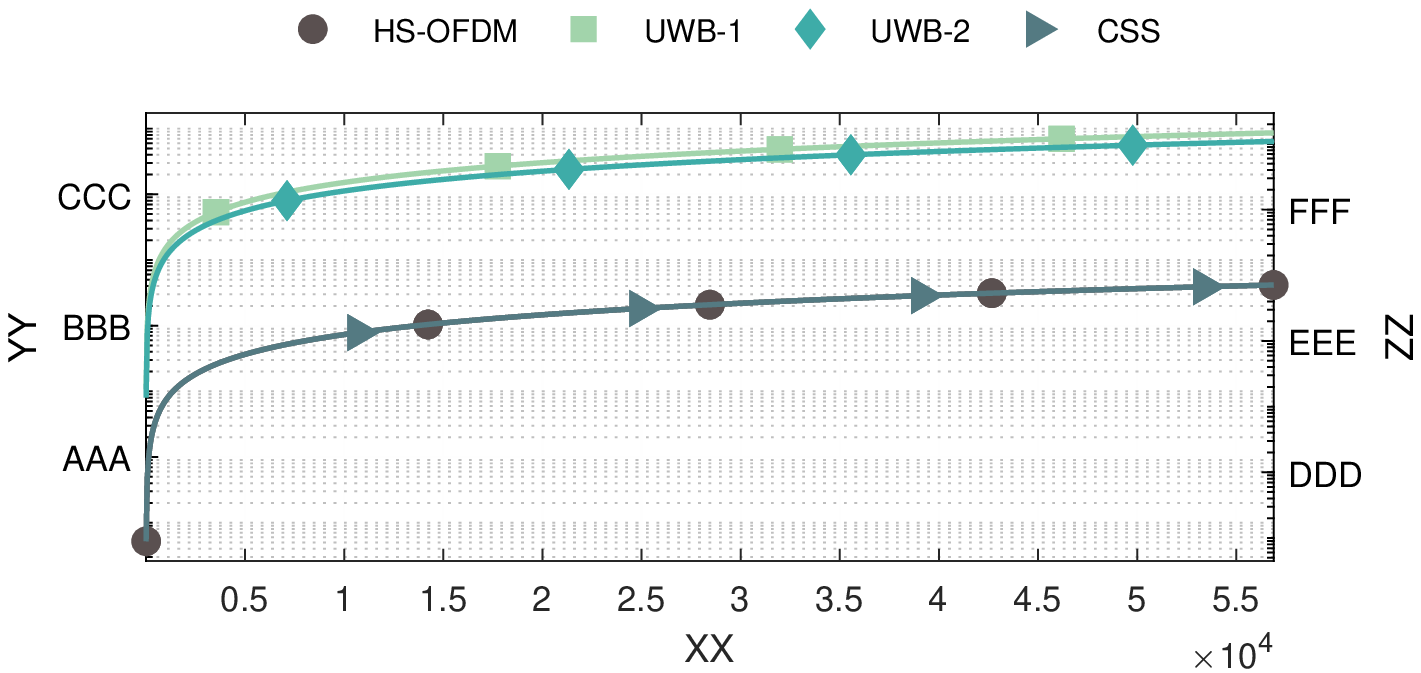}\label{fig:NB_Td_T}
	}\\
	\subfloat[ ]{
		\psfrag{XX}[c][c]{$T$($\mu$s)}
		\psfrag{YY}[c][c]{$\delta_{\text{LV}}$(km)}
		\psfrag{ZZ}[c][c]{$\delta_{\text{MV}}$(km)}
		\psfrag{AAA}[c][c]{\scriptsize 10$^{\text{-4}}$}
		\psfrag{BBB}[c][c]{\scriptsize 10$^{\text{-2}}$}
		\psfrag{CCC}[c][c]{\scriptsize 10$^{\text{0}}$}
		\psfrag{DDD}[c][c]{\scriptsize 10$^{\text{-4}}$}
		\psfrag{EEE}[c][c]{\scriptsize 10$^{\text{-2}}$}
		\psfrag{FFF}[c][c]{\scriptsize 10$^{\text{0}}$}
		\includegraphics[width=8.5cm]{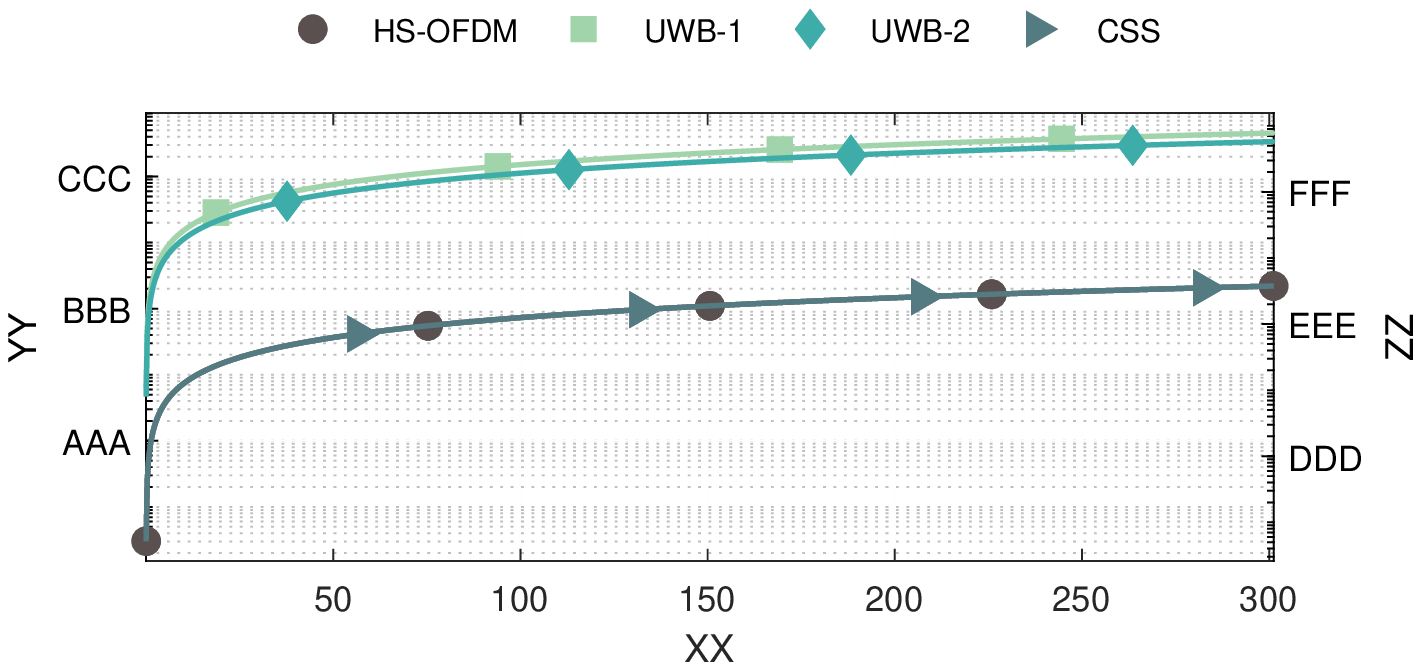}\label{fig:BB_Td_T}
	}\\
	\caption{Range resolution $\delta_{LV}$ and $\delta_{MV}$ for respectively \ac{LV} and \ac{MV} scenarios as a function of the pulse duration $T$ for (a) \ac{NB}-\ac{PLC} and (b) \ac{BB}-\ac{PLC} systems.}\label{fig:Td_T}
\end{figure}

The second aspect of our analysis is based on range resolution and \ac{PCR}, being the former shown in Figs.~\ref{fig:NB_Td_T} and \ref{fig:BB_Td_T} as a function of $T$ for all four \ac{PLC} pulses for \ac{NB}- and \ac{BB}-\ac{PLC} systems and the latter in Table~\ref{tab:PCR}, respectively. In these figures, the left vertical axes show the range resolution $\delta_{\text{LV}}$ for the considered low-voltage power distribution network, while the vertical axes on the right side shows range resolution $\delta_{\text{MV}}$ for the considered medium-voltage power distribution network.  Once again the \ac{UWB} pulses present very similar behavior, with the first and second derivatives of the Gaussian pulse presenting low \ac{PCR}, i.e., $PCR_{UWB-1} = 4.95$ and $PCR_{UWB-2} = 6.67$, respectively. In this analysis, the finest resolution, i.e., the best capability of resolving two close faults, is achieved by \ac{HS-OFDM} and \ac{CSS} pulses. With the chosen parameters, these pulses present the highest \acp{PCR}, which are $PCR_{HS-OFDM} = PCR_{CSS} = 1024$. For \ac{NB}, the achieved range resolution values range from a few meters for short pulses to tens of thousands of kilometers for longer pulses, with a ratio of $1.71$ between the resolution in the \ac{MV} and \ac{LV} and  scenarios due to their different phase velocity. \ac{BB} pulses, on the other hand, yield range resolution values from a few millimeters for short pulses to a few meters for longer pulses with the same ratio between \ac{MV} and \ac{LV} cases.
As \ac{PLC} signaling in \ac{LV} and \ac{MV} power distribution networks may cover lengths of hundreds of meters \cite{PLCbook2016} and units to tens of kilometers \cite{1901_2}, respectively, the presented results indicate that \ac{NB} pulses provide a fair capability of resolving close impedance discontinuities in most cases, being the most suitable option for most distribution network scenarios. The use of \ac{BB} pulses is therefore limited to cases where a very fine resolution is desired, such as very short sections of distribution networks.

\begin{table}[!b]
	\footnotesize
	\centering
	\caption{PCR for the considered PLC pulses.}
	\label{tab:PCR}
	\begin{tabular}{ll}
		\hline\hline
		\multicolumn{1}{c}{Pulse} & \multicolumn{1}{c}{PCR}\\ \hline\hline
		\multicolumn{1}{c}{HS-OFDM}  & \multicolumn{1}{c}{$1024$} \\
		\multicolumn{1}{c}{UWB-1}    & \multicolumn{1}{c}{$4.95$}  \\
		\multicolumn{1}{c}{UWB-2}    & \multicolumn{1}{c}{$6.67$}  \\
		\multicolumn{1}{c}{CSS}      & \multicolumn{1}{c}{$1024$} \\ \hline\hline              
	\end{tabular}
\end{table}

A further topic of our analysis is the comparison among the four considered \ac{PLC} pulses in terms of \ac{PSLR} and \ac{ISLR}, whose values are shown in Table~\ref{tab:metrics}. In this table, we see that \ac{UWB} pulses present the worst performances in terms of \ac{PSLR}, being outperformed by the \ac{HS-OFDM} pulse. The \ac{CSS} pulse yields the best \ac{PSLR} value, which is somewhat lower than the $13.2$~dB reported in the literature for bandpass chirps in radar applications. In terms of \ac{ISLR}, however, the performance ranking is altered. The first derivative of the Gaussian pulse presents the best \ac{ISLR}, followed by \ac{CSS} pulse, the second derivative of the Gaussian pulse and the \ac{HS-OFDM} pulse, in this order. Such result is explained by the fact that, disregarding the sidelobes surrounding the main lobe, $R_{pp,UWB-1}(\tau)$ presents insignificant sidelobe level, as seen in Figure~\ref{fig:Rp2}. This, however, does not hold for $R_{pp,UWB-2}(\tau)$, as seen in Figure~\ref{fig:Rp3}. The relevant \ac{ISLR} for the \ac{HS-OFDM} and \ac{CSS} pulses in comparison to the obtained value for the second derivative of the Gaussian pulse is explained by their higher \ac{PCR}. As a matter of fact, the sidelobe level of both $R_{pp,OFDM}(\tau)$ and $R_{pp,CS}(\tau)$ is not negligible even at the extremities and their main lobes are very narrow and, as a consequence, the ratio between them yields an expressive value. We finally emphasize that, in the \ac{HS-OFDM} case, better results may be achieved via appropriate pulse design with more robust techniques such as in \cite{mohseni2010,lellouch2016}.

\begin{table}[!b]
	\footnotesize
	\centering
	\caption{ISLR and PSLR for the considered PLC pulses.}
	\label{tab:metrics}
	\begin{tabular}{lll}
		\hline\hline
		\multicolumn{1}{c}{Pulse} & \multicolumn{1}{c}{PSLR} & \multicolumn{1}{c}{ISLR}\\ \hline\hline
		\multicolumn{1}{c}{HS-OFDM}  & \multicolumn{1}{c}{$-13.42$dB} & \multicolumn{1}{c}{$-0.67$dB} \\
		\multicolumn{1}{c}{UWB-1}    & \multicolumn{1}{c}{$-7.01$dB}  & \multicolumn{1}{c}{$-3.77$dB} \\
		\multicolumn{1}{c}{UWB-2}    & \multicolumn{1}{c}{$-4.18$dB}  & \multicolumn{1}{c}{$-0.94$dB} \\
		\multicolumn{1}{c}{CSS}      & \multicolumn{1}{c}{$-14.04$dB} & \multicolumn{1}{c}{$-1.75$dB} \\ \hline\hline              
	\end{tabular}
\end{table}

\begin{figure}[!t]
	\centering
	\subfloat[ ]{
		\psfrag{XX}[c][c]{$T$($\mu$s)}
		\psfrag{YY}[c][c]{$d_{\max,\text{LV}}$(km)}
		\psfrag{ZZ}[c][c]{$d_{\max,\text{MV}}$(km)}
		\includegraphics[width=8.5cm]{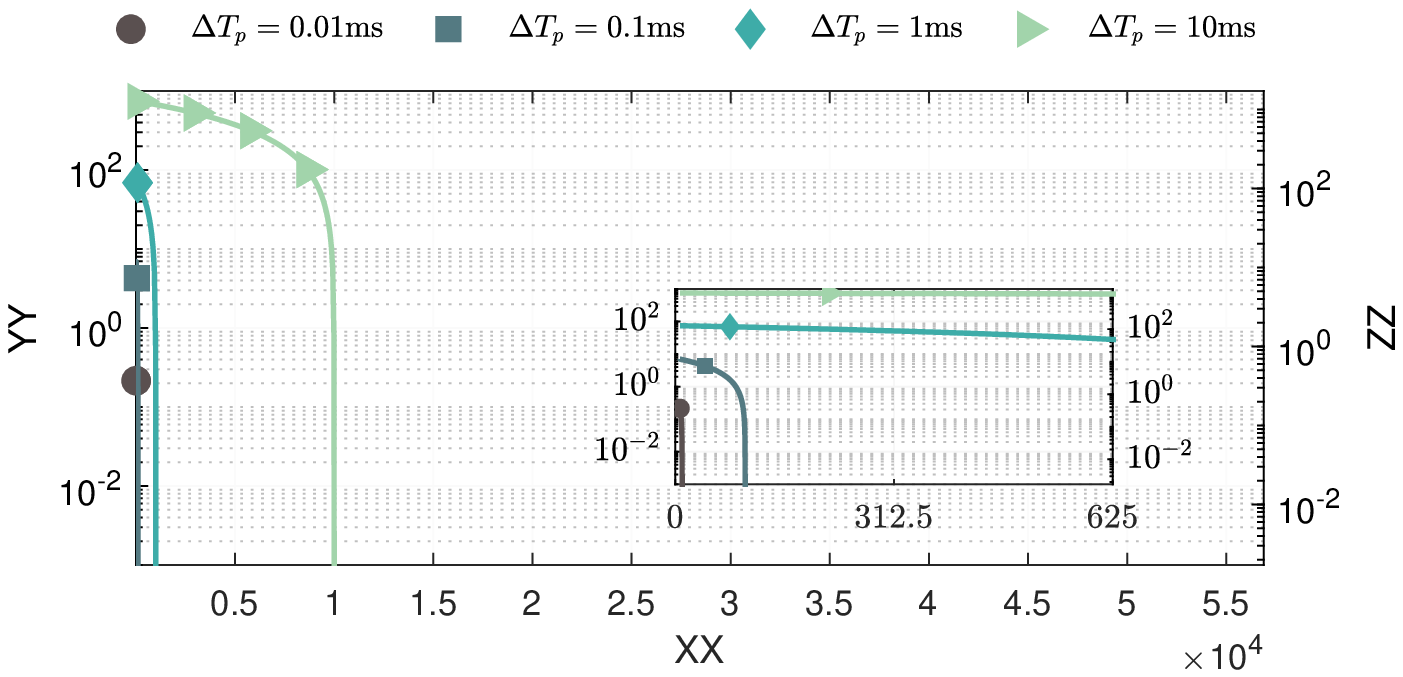}\label{fig:MUR_NB}
	}\\
	\subfloat[ ]{
		\psfrag{XX}[c][c]{$T$($\mu$s)}
		\psfrag{YY}[c][c]{$d_{\max,\text{LV}}$(km)}
		\psfrag{ZZ}[c][c]{$d_{\max,\text{MV}}$(km)}
		\includegraphics[width=8.5cm]{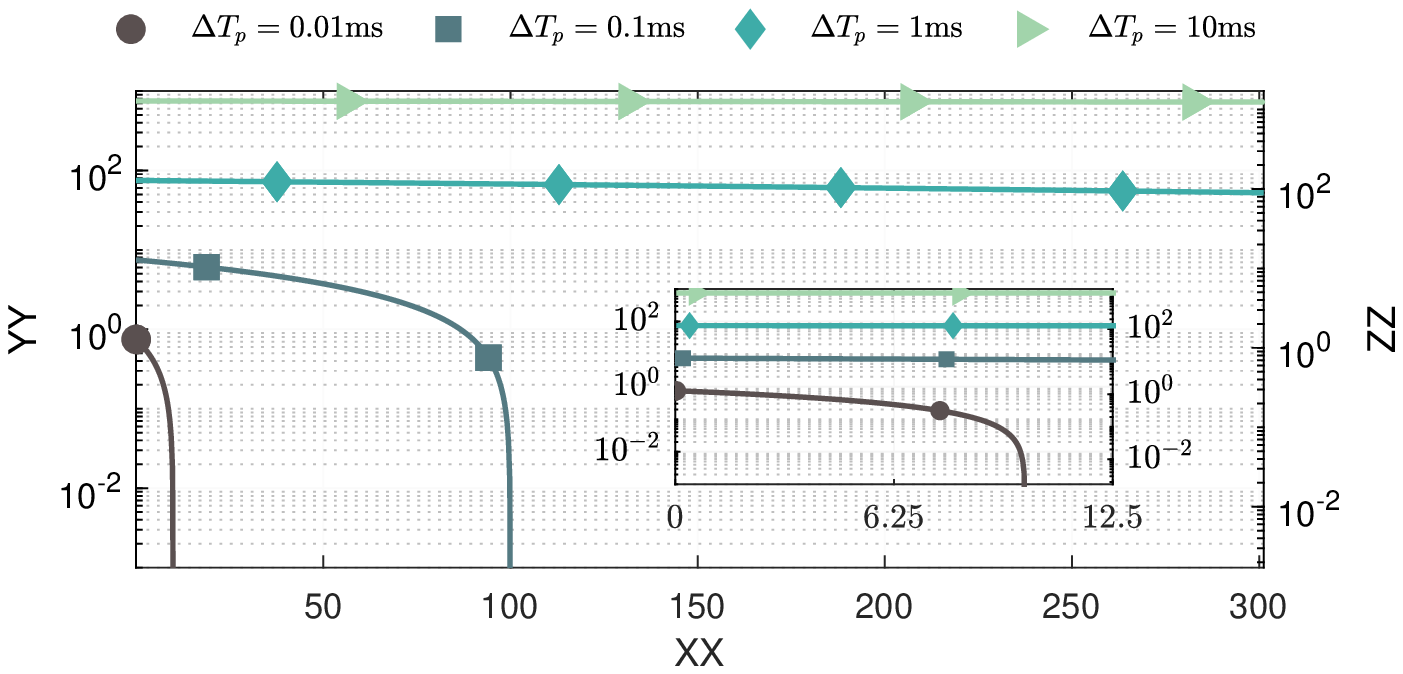}\label{fig:MUR_BB}
	}\\
	\caption{Maximum unambiguous range $d_{\max,\text{LV}}$ and $d_{\max,\text{MV}}$ for respectively \ac{LV} and \ac{MV} scenarios as a function of the pulse duration $T$ for (a) \ac{NB}-\ac{PLC} and (b) \ac{BB}-\ac{PLC} systems.}\label{fig:MUR}
\end{figure}

We next turn our attention to the maximum unambiguous ranges achieved by the four \ac{PLC} pulses, which are shown in Figs.~\ref{fig:MUR_NB} and \ref{fig:MUR_BB} for \ac{NB}- and \ac{BB}-\ac{PLC}, respectively. In these figures, the left vertical axis shows $d_{\max,\text{LV}}$ for the considered low-voltage power distribution network, whereas the vertical axis on the right side shows $d_{\max,\text{MV}}$ for the considered medium-voltage power distribution network. We adopted pulse repetition intervals of $\Delta T_p=0.01$ms, $\Delta T_p=0.1$ms, $\Delta T_p=1$ms, and $\Delta T_p=10$ms. In these figures, we see that longer pulses result in a shorter maximum unambiguous range for a given pulse interval rate, as one can easily conclude by analyzing \eqref{eq:max_range}. In this sense, \ac{NB} pulses present shorter maximum unambiguous range values than \ac{BB} pulses due to their longer duration, which is clear through a comparison of Figures \ref{fig:MUR_NB} and \ref{fig:MUR_BB}. We also see that pulse repetition intervals of $\Delta T_p=1$ms and $\Delta T_p=0.01$ms cover typical \ac{MV} and \ac{LV} distances for \ac{NB} pulses for $T$ values associated with great part of the \ac{NB} frequency range, while $\Delta T_p$ of a few unities of milliseconds are necessary to cover both \ac{MV} and \ac{LV} scenarios for \ac{BB} pulses.

\subsection*{Regulatory compliance}\label{subsec:res_reg}

Although proper resolution and maximum unambiguous range can be achieved through an efficient project of the pulse to be injected into the distribution network, regulatory constraints reduce the degree of freedom of this project and therefore limits the set of possible $T$, $\delta$, and $d_{\max}$ values. In this sense, we carry out an analysis covering the \ac{NB} regulations \ac{CENELEC}, \ac{ARIB}, and \ac{FCC}, whereas we address the European and Brazilian regulations \cite{girotto2017,ANATEL2009} and the HomePlug AV2 extended frequency band \cite{homeplugAV2} for the \ac{BB} case. In this analysis, we consider frequency bands from zero to the frequency upper bound of each regulation, achieving $B=148.50$~kHz, $B=450$~kHz, and $B=490$~kHz for \ac{CENELEC}, \ac{ARIB}, and \ac{FCC}, and $B=30$~MHz, $B=50$~MHz, and $B=86$~MHz for the European and Brazilian regulations, and the HomePlug AV2 specification.
Based on these considerations, Table~\ref{tab:reg_T} presents the pulse duration for each of the considered regulations. As expected from the analysis from Figure~\ref{fig:B_T}, pulses associated with shorter frequency bandwidths present longer duration. 

The resulting range resolution $\delta$ for the considered pulses in the considered \ac{LV} and \ac{MV} scenarios are presented in Table~\ref{tab:reg_RR}. The attained results reinforce the suitability of \ac{BB} pulses only for very short sections of distribution networks. In spite of providing fine resolution, these pulses occupy rather long frequency bandwidths that tend to result in larger attenuation of the injected signal, which becomes unpractical in long \ac{LV} distribution networks or \ac{MV} distribution networks in general. \ac{NB} pulses that comply with the considered regulations, in their turn, are appropriate for most scenarios, with the \ac{HS-OFDM} and \ac{CSS} pulses in the \ac{FCC} band offering the finest resolution in both \ac{LV} and \ac{MV} scenarios. \ac{UWB} pulses, however, must be used with care due to their great level of sidelobes.

\begin{table}[!b]
	\footnotesize
	\centering
	\caption{Pulse duration $T$ associated with regulations for \ac{NB}- and \ac{BB}-\ac{PLC} systems.}
	\label{tab:reg_T}
	\begin{tabular}{ccccc}
		\hline\hline
		\multicolumn{5}{c}{$T$($\mu$s)}\\ \hline\hline
		\multicolumn{1}{l}{}                         & \multicolumn{1}{l|}{}        & \multicolumn{1}{c|}{CENELEC}  & \multicolumn{1}{c|}{ARIB}      & FCC    \\ \hline
		\multicolumn{1}{c|}{\multirow{4}{*}{NB-PLC}} & \multicolumn{1}{c|}{HS-OFDM} & \multicolumn{1}{c|}{3447.81}     & \multicolumn{1}{c|}{1137.78}      & 1044.90   \\
		\multicolumn{1}{c|}{}                        & \multicolumn{1}{c|}{UWB-1}   & \multicolumn{1}{c|}{24.00}     & \multicolumn{1}{c|}{7.92}      & 7.27   \\
		\multicolumn{1}{c|}{}                        & \multicolumn{1}{c|}{UWB-2}   & \multicolumn{1}{c|}{26.62}     & \multicolumn{1}{c|}{8.78}      & 8.07   \\
		\multicolumn{1}{c|}{}                        & \multicolumn{1}{c|}{CSS}     & \multicolumn{1}{c|}{3447.81}     & \multicolumn{1}{c|}{1137.78}      & 1044.90   \\ \hline\hline
		& \multicolumn{1}{l|}{}        & \multicolumn{1}{c|}{European} & \multicolumn{1}{c|}{Brazilian} & HomePlug \\ \hline
		\multicolumn{1}{c|}{\multirow{4}{*}{BB-PLC}} & \multicolumn{1}{c|}{HS-OFDM} & \multicolumn{1}{c|}{17.07}     & \multicolumn{1}{c|}{10.24}      & 5.95   \\
		\multicolumn{1}{c|}{}                        & \multicolumn{1}{c|}{UWB-1}   & \multicolumn{1}{c|}{0.12}     & \multicolumn{1}{c|}{0.07}      & 0.04   \\
		\multicolumn{1}{c|}{}                        & \multicolumn{1}{c|}{UWB-2}   & \multicolumn{1}{c|}{0.13}     & \multicolumn{1}{c|}{0.08}      & 0.05   \\
		\multicolumn{1}{c|}{}                        & \multicolumn{1}{c|}{CSS}     & \multicolumn{1}{c|}{17.07}     & \multicolumn{1}{c|}{10.24}      & 5.95   \\ \hline
	\end{tabular}
\end{table}

\begin{table}[!t]
	\footnotesize
	\centering
	\caption{Range resolution $\delta$ associated with regulations for \ac{NB}- and \ac{BB}-\ac{PLC} systems.}
	\label{tab:reg_RR}
	\begin{tabular}{cccccc}
		\hline\hline
		\multicolumn{6}{c}{$\delta$(m)}\\ \hline\hline
		\multicolumn{1}{l}{}                         & \multicolumn{1}{l}{}                     & \multicolumn{1}{l|}{}        & \multicolumn{1}{c|}{CENELEC}  & \multicolumn{1}{c|}{ARIB}      & FCC    \\ \hline
		\multicolumn{1}{c|}{\multirow{8}{*}{NB-PLC}} & \multicolumn{1}{c|}{\multirow{4}{*}{LV}} & \multicolumn{1}{c|}{HS-OFDM} & \multicolumn{1}{c|}{504.71}   & \multicolumn{1}{c|}{166.56}    & 152.96 \\
		\multicolumn{1}{c|}{}                        & \multicolumn{1}{c|}{}                    & \multicolumn{1}{c|}{UWB-1}   & \multicolumn{1}{c|}{726.81}   & \multicolumn{1}{c|}{239.85}    & 220.27 \\
		\multicolumn{1}{c|}{}                        & \multicolumn{1}{c|}{}                    & \multicolumn{1}{c|}{UWB-2}   & \multicolumn{1}{c|}{598.07}   & \multicolumn{1}{c|}{197.36}    & 181.25 \\
		\multicolumn{1}{c|}{}                        & \multicolumn{1}{c|}{}                    & \multicolumn{1}{c|}{CSS}     & \multicolumn{1}{c|}{504.71}   & \multicolumn{1}{c|}{166.56}    & 152.96 \\ \cline{2-6} 
		\multicolumn{1}{c|}{}                        & \multicolumn{1}{c|}{\multirow{4}{*}{MV}} & \multicolumn{1}{c|}{HS-OFDM} & \multicolumn{1}{c|}{863.60}   & \multicolumn{1}{c|}{284.99}    & 261.72 \\
		\multicolumn{1}{c|}{}                        & \multicolumn{1}{c|}{}                    & \multicolumn{1}{c|}{UWB-1}   & \multicolumn{1}{c|}{1243.63}   & \multicolumn{1}{c|}{410.40}    & 376.90 \\
		\multicolumn{1}{c|}{}                        & \multicolumn{1}{c|}{}                    & \multicolumn{1}{c|}{UWB-2}   & \multicolumn{1}{c|}{1023.34}   & \multicolumn{1}{c|}{337.70}    & 310.13 \\
		\multicolumn{1}{c|}{}                        & \multicolumn{1}{c|}{}                    & \multicolumn{1}{c|}{CSS}     & \multicolumn{1}{c|}{863.60}   & \multicolumn{1}{c|}{284.99}    & 261.72 \\ \hline\hline
		\multicolumn{1}{l}{}                         & \multicolumn{1}{l}{}                     & \multicolumn{1}{l|}{}        & \multicolumn{1}{c|}{European} & \multicolumn{1}{c|}{Brazilian} & HomePlug \\ \hline
		\multicolumn{1}{c|}{\multirow{8}{*}{BB-PLC}} & \multicolumn{1}{c|}{\multirow{4}{*}{LV}} & \multicolumn{1}{c|}{HS-OFDM} & \multicolumn{1}{c|}{2.50}   & \multicolumn{1}{c|}{1.50}    & 0.87 \\
		\multicolumn{1}{c|}{}                        & \multicolumn{1}{c|}{}                    & \multicolumn{1}{c|}{UWB-1}   & \multicolumn{1}{c|}{3.60}   & \multicolumn{1}{c|}{2.16}    & 1.25 \\
		\multicolumn{1}{c|}{}                        & \multicolumn{1}{c|}{}                    & \multicolumn{1}{c|}{UWB-2}   & \multicolumn{1}{c|}{2.96}   & \multicolumn{1}{c|}{1.78}    & 1.03 \\
		\multicolumn{1}{c|}{}                        & \multicolumn{1}{c|}{}                    & \multicolumn{1}{c|}{CSS}     & \multicolumn{1}{c|}{2.50}   & \multicolumn{1}{c|}{1.50}    & 0.87 \\ \cline{2-6} 
		\multicolumn{1}{c|}{}                        & \multicolumn{1}{c|}{\multirow{4}{*}{MV}} & \multicolumn{1}{c|}{HS-OFDM} & \multicolumn{1}{c|}{4.27}   & \multicolumn{1}{c|}{2.56}    & 1.49 \\
		\multicolumn{1}{c|}{}                        & \multicolumn{1}{c|}{}                    & \multicolumn{1}{c|}{UWB-1}   & \multicolumn{1}{c|}{6.16}   & \multicolumn{1}{c|}{3.69}    & 2.15 \\
		\multicolumn{1}{c|}{}                        & \multicolumn{1}{c|}{}                    & \multicolumn{1}{c|}{UWB-2}   & \multicolumn{1}{c|}{5.07}   & \multicolumn{1}{c|}{3.04}    & 1.77 \\
		\multicolumn{1}{c|}{}                        & \multicolumn{1}{c|}{}                    & \multicolumn{1}{c|}{CSS}     & \multicolumn{1}{c|}{4.27}   & \multicolumn{1}{c|}{2.56}    & 1.49 \\ \hline
	\end{tabular}
\end{table}

Assuming a desired maximum unambiguous range $d_{\max}=~1$~km for \ac{LV} distribution networks and $d_{\max}=10$~km for \ac{LV} distribution networks, Table~\ref{tab:reg_MUR} shows the necessary pulse repetition interval $\Delta T$ for the analyzed \ac{PLC} pulses with durations presented in Table~\ref{tab:reg_T}. As we consider same duration $T$ for the \ac{HS-OFDM} and \ac{CSS} pulses, they require same pulse repetition interval. Such rate is considerably longer than the ones required for \ac{UWB} pulses, with the difference between them decreasing as the frequency bandwidth $B$ increases. This means that, although \ac{UWB} pulses offer a worse performance in terms of \ac{PSLR}, \ac{ISLR}, and even range resolution $\delta$, their adoption allows the acquisition of a very larger number of reflectograms in a given time interval than if \ac{HS-OFDM} or \ac{CSS} pulses are adopted.

\section{Conclusion}\label{sec:conclusion}

This work has discussed key aspects of typical \ac{PLC} pulses for fault sensing in power lines via pulse-compression \ac{TDR}. In summary, the fault sensing system has been described, closed-form expressions for the compressed \ac{PLC} pulses have been presented, and performance metrics usually adopted for radar systems have been used for assessing the suitability of the analyzed signals for \ac{TDR} purposes.

Moreover, we have presented numerical results for two scenarios, namely an European underground \ac{LV} distribution network and an US overhead \ac{MV} distribution network, carrying out a comparative analysis among \ac{HS-OFDM}, impulsive \ac{UWB}, and \ac{CSS} pulses of same time duration in terms of occupied frequency bandwidth, range resolution, and pulse repetition interval. Finally, we concluded our discussion addressing \ac{NB} and \ac{BB} \ac{PLC} regulatory compliance.

Based on the obtained results, we see that the use of impulsive \ac{UWB} pulses, i.e., the first and the second derivatives of the Gaussian pulse, enables the obtaining of a higher number of reflectograms with finer resolution due to their short time duration. However, the pulse compression for this class of signals is not nearly as effective as for \ac{HS-OFDM} and \ac{CSS} pulses, resulting therefore in poorer range resolution enhancement.

It has also been revealed that the best \ac{PSLR} value is achieved by \ac{CSS} pulses, followed by \ac{HS-OFDM}, first derivative of the Gaussian pulse, and second derivative of the Gaussian pulse, in this order. With respect to \ac{ISLR}, the best performance is achieved by \ac{HS-OFDM} pulses, followed by the second derivative of the Gaussian pulse, \ac{CSS} pulses and the first derivative of the Gaussian pulse, in this order. Based on the \ac{PSLR} values, we can conclude that the use of \ac{HS-OFDM} and \ac{CSS} pulses results in reflectograms with less distortion. The relevant \ac{ISLR} value for \ac{HS-OFDM} and \ac{CSS} pulses in comparison with the second derivative of the Gaussian pulse is explained by the higher \ac{PCR} achieved by the former, which makes the integrated sidelobe level higher although the sidelobes have, individually, lower level than in \ac{UWB} pulses.

\begin{table}[!t]
	\footnotesize
	\centering
	\caption{Pulse repetition interval $\Delta T$ associated with $d_{\max}~=~1$~km for the \ac{LV} scenario and $d_{\max}=10$~km for the \ac{MV} scenario complying with regulations for \ac{NB}- and \ac{BB}-\ac{PLC} systems.}
	\label{tab:reg_MUR}
	\begin{tabular}{cccccc}
		\hline\hline
		\multicolumn{6}{c}{$\Delta T$($\mu$s)}\\ \hline\hline
		\multicolumn{1}{l}{}                         & \multicolumn{1}{l}{}                     & \multicolumn{1}{l|}{}        & \multicolumn{1}{c|}{CENELEC}  & \multicolumn{1}{c|}{ARIB}      & FCC    \\ \hline
		\multicolumn{1}{c|}{\multirow{8}{*}{NB-PLC}} & \multicolumn{1}{c|}{\multirow{4}{*}{LV}} & \multicolumn{1}{c|}{HS-OFDM} & \multicolumn{1}{c|}{3461.15}   & \multicolumn{1}{c|}{1151.12}    & 1058.24 \\
		\multicolumn{1}{c|}{}                        & \multicolumn{1}{c|}{}                    & \multicolumn{1}{c|}{UWB-1}   & \multicolumn{1}{c|}{37.34}   & \multicolumn{1}{c|}{21.26}    & 20.62 \\
		\multicolumn{1}{c|}{}                        & \multicolumn{1}{c|}{}                    & \multicolumn{1}{c|}{UWB-2}   & \multicolumn{1}{c|}{39.96}   & \multicolumn{1}{c|}{22.13}    & 21.41 \\
		\multicolumn{1}{c|}{}                        & \multicolumn{1}{c|}{}                    & \multicolumn{1}{c|}{CSS}     & \multicolumn{1}{c|}{3461.15}   & \multicolumn{1}{c|}{1151.12}    & 1058.24 \\ \cline{2-6} 
		\multicolumn{1}{c|}{}                        & \multicolumn{1}{c|}{\multirow{4}{*}{MV}} & \multicolumn{1}{c|}{HS-OFDM} & \multicolumn{1}{c|}{3525.79}   & \multicolumn{1}{c|}{1215.75}    & 1122.87 \\
		\multicolumn{1}{c|}{}                        & \multicolumn{1}{c|}{}                    & \multicolumn{1}{c|}{UWB-1}   & \multicolumn{1}{c|}{101.98}   & \multicolumn{1}{c|}{85.90}    & 85.25 \\
		\multicolumn{1}{c|}{}                        & \multicolumn{1}{c|}{}                    & \multicolumn{1}{c|}{UWB-2}   & \multicolumn{1}{c|}{104.59}   & \multicolumn{1}{c|}{86.76}    & 86.04 \\
		\multicolumn{1}{c|}{}                        & \multicolumn{1}{c|}{}                    & \multicolumn{1}{c|}{CSS}     & \multicolumn{1}{c|}{3525.79}   & \multicolumn{1}{c|}{1215.75}    & 1122.87 \\ \hline\hline
		\multicolumn{1}{l}{}                         & \multicolumn{1}{l}{}                     & \multicolumn{1}{l|}{}        & \multicolumn{1}{c|}{European} & \multicolumn{1}{c|}{Brazilian} & HomePlug \\ \hline
		\multicolumn{1}{c|}{\multirow{8}{*}{BB-PLC}} & \multicolumn{1}{c|}{\multirow{4}{*}{LV}} & \multicolumn{1}{c|}{HS-OFDM} & \multicolumn{1}{c|}{30.41}   & \multicolumn{1}{c|}{23.58}    & 19.30 \\
		\multicolumn{1}{c|}{}                        & \multicolumn{1}{c|}{}                    & \multicolumn{1}{c|}{UWB-1}   & \multicolumn{1}{c|}{13.46}   & \multicolumn{1}{c|}{13.41}    & 13.38 \\
		\multicolumn{1}{c|}{}                        & \multicolumn{1}{c|}{}                    & \multicolumn{1}{c|}{UWB-2}   & \multicolumn{1}{c|}{13.47}   & \multicolumn{1}{c|}{13.42}    & 13.39 \\
		\multicolumn{1}{c|}{}                        & \multicolumn{1}{c|}{}                    & \multicolumn{1}{c|}{CSS}     & \multicolumn{1}{c|}{30.41}   & \multicolumn{1}{c|}{23.58}    & 19.30 \\ \cline{2-6} 
		\multicolumn{1}{c|}{}                        & \multicolumn{1}{c|}{\multirow{4}{*}{MV}} & \multicolumn{1}{c|}{HS-OFDM} & \multicolumn{1}{c|}{95.04}   & \multicolumn{1}{c|}{88.22}    & 83.93 \\
		\multicolumn{1}{c|}{}                        & \multicolumn{1}{c|}{}                    & \multicolumn{1}{c|}{UWB-1}   & \multicolumn{1}{c|}{78.09}   & \multicolumn{1}{c|}{78.05}    & 78.02 \\
		\multicolumn{1}{c|}{}                        & \multicolumn{1}{c|}{}                    & \multicolumn{1}{c|}{UWB-2}   & \multicolumn{1}{c|}{78.11}   & \multicolumn{1}{c|}{78.05}    & 78.02 \\
		\multicolumn{1}{c|}{}                        & \multicolumn{1}{c|}{}                    & \multicolumn{1}{c|}{CSS}     & \multicolumn{1}{c|}{95.04}   & \multicolumn{1}{c|}{88.22}    & 83.93 \\ \hline
	\end{tabular}
\end{table}

In a frequency bandwidth perspective, the best range resolution is achieved by \ac{HS-OFDM} and \ac{CSS} pulses, followed by the second and the first derivatives of the Gaussian pulse, in this order. Regarding regulatory compliance, we conclude that \ac{PLC} pulses occupying a frequency bandwidth within the \ac{NB} frequency range tend to be suitable for fault sensing in most distribution network scenarios, with the exception of rather short sections distribution networks.

\bibliographystyle{elsarticle-num}
\bibliography{referencias}

\end{document}